\documentclass[review, 3p, number,12pt]{elsarticle}
\usepackage{graphicx,graphics}
\usepackage{syntonly}
\usepackage{amssymb,amsmath,amsfonts}
\usepackage{subfigure}
\usepackage{multirow}
\usepackage{subfigure,soul,color}
\newcommand{\bxi}{\boldsymbol \xi}
\newcommand{\boldeta}{\boldsymbol \eta}
\newcommand{\bx}{\mathbf x}
\newcommand{\bu}{\mathbf u}
\newcommand{\by}{\mathbf y}
\newcommand{\bff}{\mathbf f}
\newcommand{\ud}{\mathbf d}
\newcommand{\bB}{\mathbf \Omega}
\newcommand{\bn}{\mathbf n}

\begin{document}

\title{{Dual-horizon Peridynamics}}

\author[TJ,BUW]{Huilong Ren}
\author[TJ,BUW]{Xiaoying Zhuang}
\ead{xiaoying.zhuang@uni-weimar.de;}
\author[TJ]{Yongchang Cai}
\author[BUW,KU]{Timon Rabczuk}
\ead{timon.rabzuk@uni-weimar.de;}

\cortext[cor1]{Corresponding author: Tel: +44 191 334 2504}

\address[TJ]{State Key Laboratory of Disaster Reduction in Civil Engineering, College of Civil Engineering,Tongji University, Shanghai 200092, China}
\address[BUW]{Institute of Structural Mechanics, Bauhaus-Universit\"at Weimar, Germany }
\address[KU]{School of Civil, Environmental and Architectural Engineering, Korea University, South Korea}


\begin{abstract} 
In this paper we develop a new Peridynamic approach  that naturally includes varying horizon sizes and completely solves the ''ghost force" issue. Therefore, the concept of dual-horizon is introduced to consider the unbalanced interactions between the particles with different horizon sizes. The present formulation is proved to fulfill both the balances of linear momentum and angular momentum. Neither the "partial stress tensor" nor the "`slice" technique are needed to ameliorate the ghost force issue in \cite{Silling2014}. The consistency of reaction forces is naturally fulfilled by a unified simple formulation. The method can be easily implemented to any existing peridynamics code with minimal changes. A simple adaptive refinement  procedure is proposed minimizing the computational cost. The method is applied here to the three Peridynamic formulations, namely bond based, ordinary state based and non-ordinary state based Peridynamics. Both two- and three- dimensional examples including the Kalthof-Winkler experiment and plate with branching cracks are tested to demonstrate the capability of the method in solving wave propagation, fracture and adaptive analysis .

\end{abstract} \vspace{10pt}

\maketitle
\section{Introduction}\label{sec:introduction}
Peridynamics has recently attracted wide interests for researchers in computational solid mechanics since it provides the possibility to model dynamic fracture with ease. The crack is part of the solution from PD simulation instead of part of the problem and no representation of the crack topology is needed. The original PD method was proposed by Silling \cite{Silling2000} in 2000 and has been exploited onwards for extensive applications of mechanical problems including impact loading, fragmentation \cite{Lai2015, Gerstle2007}, composites delamination \cite{Oterkus2012}, beam and plate structures \cite{Moyer2014, OGrady2014}.

In PD, the classical balance equations are formulated in an integral form instead of partial differential form. The particles interact with each other when the distances between the particles are within a threshold value, called \textit{horizon}. The equations of motion at any time $t$ is expressed as
\begin{align}
\label{eq:oldPD}
\rho \mathbf{\ddot{u}}(\bx,t)=\int_{H_{\bx}} \bff \left(\bu(\bx',t)-\bu(\bx,t),\bx'-\bx\right)-\bff \left(\bu(\bx,t)-\bu(\bx',t),\bx-\bx'\right)  \ud V_{\bx'} +\mathbf b(\bx,t)  \ ,
\end{align}
where $H_{\bx}$ denotes the horizon (spherical domain) belonging to $\bx$, $\bu$ is the displacement vector, $\mathbf{b}$ denotes the body force, $\rho$ is mass density in the reference configuration, and $\bff$ is a pairwise force function that computes the force vector (per unit volume squared) \cite{Silling2005}. Eq.(\ref{eq:oldPD}) shows the key idea of PD to unify continuous and discontinuous media within a single consistent set of equations. The governing equation is written in an integral form instead of an partial differential form. The crack surfaces are formed as the outcome of motion and constitutive models, and there is no entanglement of additional crack kinematics or geometry treatment. The fracture behaviour including crack branching and coalescence of multiple cracks is captured through the breakage of the bonds between particles. Therefore, the need of smoothing of crack surfaces or branching criterion in the extended finite element method (XFEM) \cite{Belytschko19}, meshless methods \cite{Rabczuk5} or other partition of unity methods (PUM) \cite{Babuska-BM:95} is completely removed. The extension of PD from 2D to 3D problems is greatly facilitated for the computer implementation, which is not always the case in other methods \cite{Fries3, Gravouil1}.

The original PD started with the bond-based formulation (BB-PD) where the bonds behave like springs and independent of each other. BB-PD can be regarded as a special case of a more general theory, the state-based peridynamics (SB-PD) \cite{Silling2000, Silling2008} which can be suited for, theoretically,  any type of constitutive model and large deformation analysis. The key difference between SB-PD and BB-PD is that in the former, the bond deformation depends on collective deformation of other bonds, whereas the bonds in the latter deforms independently. The SB-PD was later  extended into two types, namely the the ordinary state based peridynamics (OSB-PD) and the non-ordinary state based peridynamics (NOSB-PD). In all the above types of PD, horizons sizes are commonly required to be constant to avoid spurious wave reflections and ghost forces between particles. However, in many applications, the spatial distribution of the particles with changing horizon sizes is necessary, e.g. adaptive refinement, multiscale modelling and multibody analysis. In other words, in order to achieve acceptable accuracy, the entire numerical model has to be discretised with respect to the highest particle resolution locally required, and the smallest horizon size used accordingly. This is computationally expensive and undesirable. The restriction of horizons being position independent practically reduces the efficiency of PD.

In this paper, we aim to remove the issue of varying horizons and ghost force by developing a new PD formulation. The new approach is based on the concept of horizon and dual-horizon. Though peridynamics has been developed for different types of physical fields, e.g. thermal field and fluid field, we confine the present work to solving solid mechanics problems. The content of the paper is outlined as follows. \S 2 begins with stating the phenomenon of the ghost forces. In \S 3, the horizon and dual-horizon are introduced. New motion equations with varying horizons are derived based on the dual-horizon concept. The balance of linear momentum and the balance of angular momentum of the present PD formulation are proved. The applications of the new formulation for BB-PD, OSB-PD and NOSB-PD are described in details in \S 4. In \S 5, three numerical examples are presented to validate the present method.

\section{Ghost force and spurious wave reflection in peridynamics}
\begin{figure}[htp]
	\centering
		\includegraphics[width=9cm]{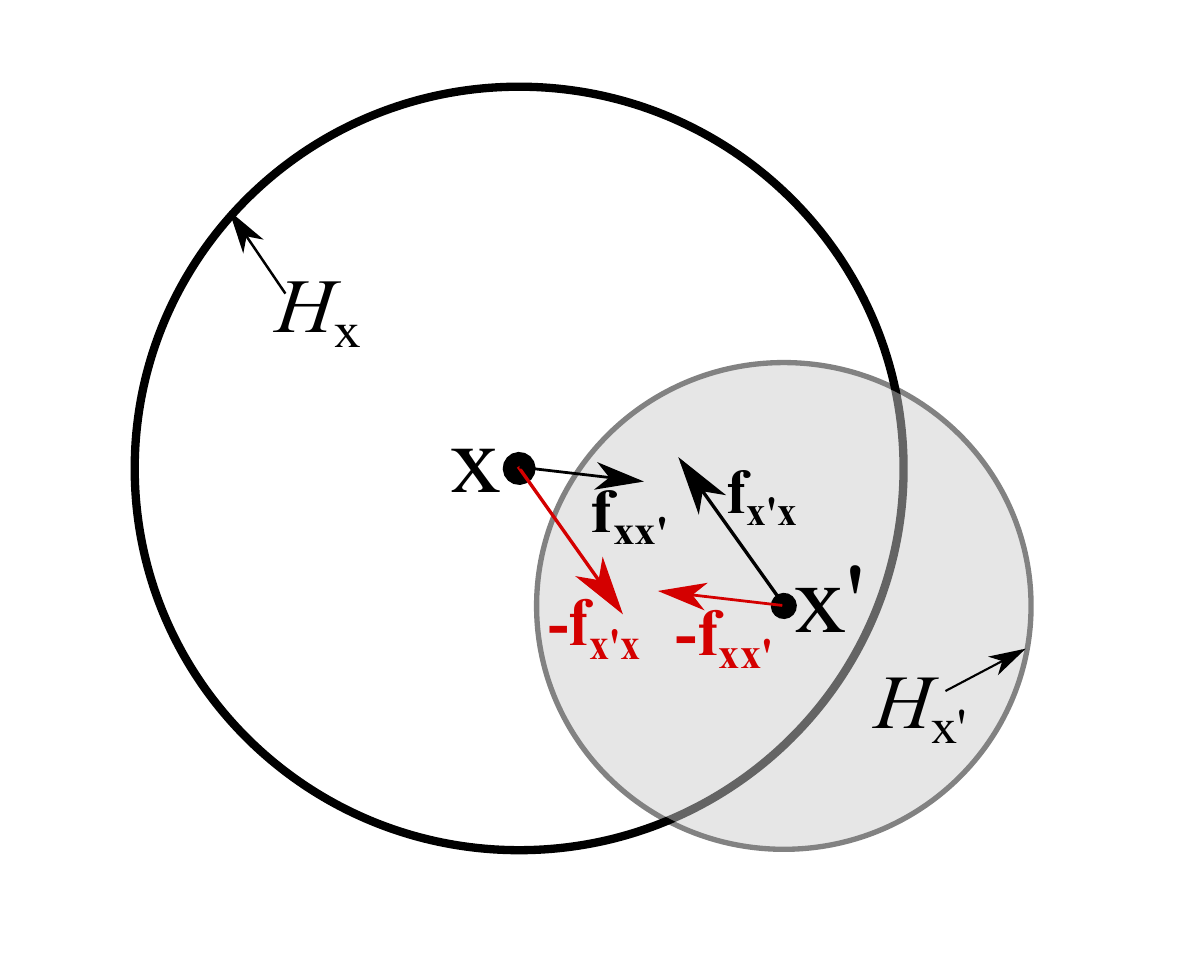}
	\caption{Force vector in peridynamics with varying horizons. Reactive force $-\bff_{\bx' \bx}$ on $\bx$ due to $\bx$ exerted on $\bx'$ as $\bx'\in H_\bx$.}
\label{fig:HorizonIJ}
\end{figure}

In the original PD theory, the forces exerted on a particle is the summation of all the pairwise forces from the particles falling inside the horizon of that particle. When the horizon sizes are set constant for all particles, the bond force is always pairwise. On the other hand, it should be noted that the size of the particles that represent the mass quantity can vary \cite{Silling2000, Silling2007}. Spurious wave reflections emerge when the horizon sizes vary. We illustrate the mechanism of the phenomenon as follows. Consider a particle $\bx'$ falling inside the horizon of particle $\bx$, see Fig.\ref{fig:HorizonIJ}. Let $\bff_{\bx \bx'}$ denote the force vector acting on particle $\bx$ due to particle $\bx'$, where the first subscript $\bx$ indicates $\bx$ being the object of force and the second subscript $\bx'$ indicates $\bff_{\bx \bx'} $ being the source of force from $\bx'$. Take the Fig. \ref{fig:HorizonIJ} for an example, as $\bx' \in H_{\bx}$, the force $\bff_{\bx' \bx}\ne 0 $, $\bx \notin H_{\bx'}$ due to unequal size of horizons, hence $\bff_{\bx \bx'}= 0$. When computing the reactive forces for particle $\bx$, the bond forces $\mathbf F_{\bx \bx'}= \bff_{\bx \bx'}-\bff_{\bx' \bx}=-\bff_{\bx' \bx} $, which is added to $\mathbf F_\bx $. Likewise, when computing forces for particle $\bx'$, particle $\bx$ exerts no force on $\bx'$ as $\bx$ is not inside the horizon of $\bx'$. Consequently, the bond force $\mathbf F_{\bx \bx'} $ only exists unilaterally, which is known as the ``ghost force" \cite{Silling2014} resulting in an unbalanced internal force. The balance of linear momentum and balance of angular momentum in this case are violated, and hence yields spurious wave reflections in PD simulations.

Efforts have been made by researchers to explore the possibility of making PD suitable for nonuniform spatial discretisation.  Bobaru et. al studied the convergence and adaptive refinement in 1D peridynamics \cite{Bobaru2009} and multi-scale modeling in 2D BB-PD \cite{Bobaru2011}.  Dipasquale et al. recently \cite{Dipasquale2014} introduced a trigger based on the damage state of the material for 2D refinement of BB-PD. Refined PD was developed in \cite{Askari2008, Bobaru2009, Bobaru2011, Dipasquale2014, Yu2011}, however it is restricted to BB-PD formulation and to the authors knowledge, has only been used for one- and two-dimensional problems. In these works, the spurious wave reflection or ghost force problem is not solved, and the refinement is performed by checking the spurious reflection is within an acceptable range compared to the magnitude of the whole wave. Recently, the \textit{partial stress} was proposed in \cite{Silling2014} to remove the ghost force for varying horizons. However, the method imposes certain restrictions to the deformation of the body and requires the computation of partial stress which is complicated. It to certain extent impairs the simplicity of the original PD, especially for BB-PD and OSB-PD. Besides, the method does not completely remove the ghost force but with a small residual which is believed to be acceptable. Though the slice method was devised in the same work, it is applicable only to piecewise constant sizes of horizons, and requires additional computation to enforce the consistency between particles close to the interface.

\section{Governing equations based on horizon and dual-horizon}\label{sec:Dualhorizons}
In this section, the concept of \textit{horizon} and \textit{dual-horizon} will be introduced to formulate the balance equations for particles with varying horizons. It is applied to all peridynamic formulations.

\subsection{\textit{Horizon} and \textit{dual-horizon}}\label{sec:HorizonConcept}
\begin{figure}[htp]
	\centering
		\includegraphics[width=8cm]{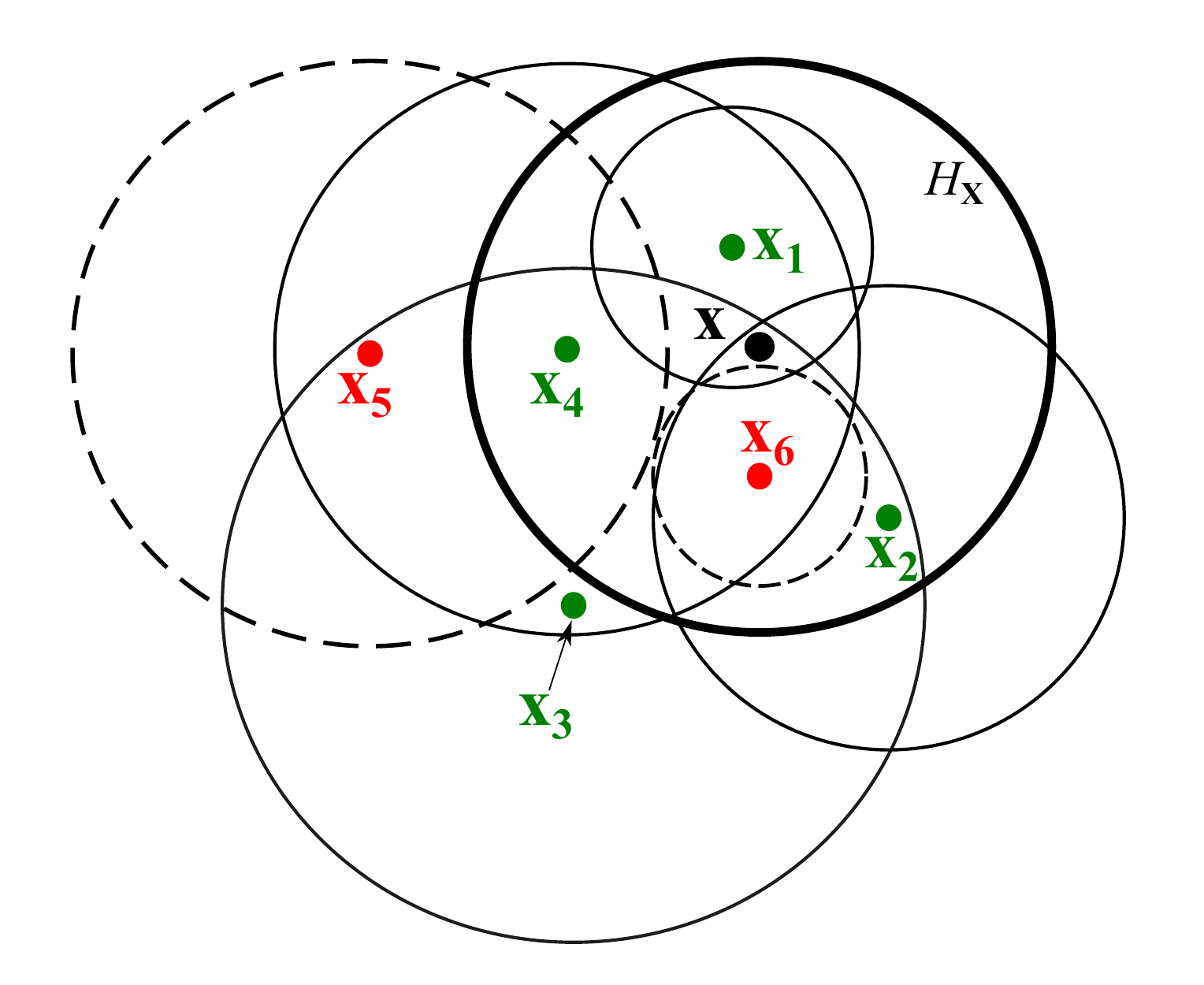}
\caption{The schematic diagram for horizon and dual-horizon, all circles above are horizons. The green points $\{\bx_1,\bx_2,\bx_3,\bx_4\} \in H_\bx'$,whose horizons denote by thin solid line; the red points $\{\bx_5,\bx_6\} \notin H_\bx'$ ,whose horizons denote by dashed line}\label{fig:4Horizon}
\end{figure}

We will begin by restating the original concept of horizon in peridynamics. In peridynamics theory, the particles interact with each other within a finite distance. A particle is considered to have an influence over other particles within a small domain centering that particle. The radius of the domain is known as ``horizon", see Fig.\ref{fig:4Horizon}. Particle $\bx$ (thick solid line) is included in the horizons of particles $\bx_1$, $\bx_2$, $\bx_3$ and $\bx_4$ (thin solid line), however is not included in the horizons of particles $\bx_5$ and $\bx_6$ (dashed line). The concept of horizon can be compared to the concept of ``nodal support" in meshless methods.

\noindent \textbf{Horizon}\\
The \textit{horizon} $H_{\bx}$ is the domain where any particle falling inside will receive the forces exerted by $\bx$. Hence, $\bx$ will undertake all the reactive or passive forces from particles in $H_{\bx}$. Hence, the horizon can be viewed as passive force horizon. The reactive force acting on $\bx $ follows Newton's third law. As shown in Fig.\ref{fig:HorizonIJ}, the reactive force $-\bff_{\bx' \bx}$ exerted on $\bx$ by other particles is in opposite direction of the forces applied by $\bx$ to other particles.  \vspace{3mm}

\noindent \textbf{Dual-horizon}\\
\textit{Dual-horizon} is defined as a union of points whose horizons include $\bx$, denoted by $H_{\bx}'=\{\bx': \bx \in H_{\bx'}\}$. In the notation of dual-horizon $H_{\bx}'$, the superscript prime indicates ``dual", and the subscript $\bx$ denotes the object particle. It can be understood as the set of the horizons that belong to the particles who can ``see" $\bx$ in their horizons. As shown in Fig.\ref{fig:4Horizon}, the dual-horizon with respect to $\bx$ is the union of $\bx_1$, $\bx_2$, $\bx_3$ and $\bx_4$, whose horizons are denoted by thin solid circles. Particles $\bx_5$ and $\bx_6$ are not included in the dual-horizon of $\bx$ since their horizons do not include $\bx$. In this case, $\bx$ becomes the object ``observed" by the other particles. If $\bx$ is within the horizon of $\bx'$, then $\bx'$ has an active or direct effect on $\bx$, corresponding to the passive effect in horizon defined previously. Particle $\bx$ receives the active forces from other particles in the \textit{dual-horizon} of $\bx$, and in this sense it is considered as ``dual" corresponding to \textit{horizon}. For any point $\bx$, the shape of  $H_{\bx}'$ is arbitrary, and depends on the sizes and shapes of horizons as well as the locations of the particles. Note that the horizon can take other shapes other than circles or spheres.

The bond inside the dual-horizon ($H_{\bx}'$) is termed as ``dual-bond", and this is corresponding to the bond in the horizon ($H_{\bx}$). It can be seen that the bond of one particle can become the dual-bond for another particle interacting with it. Note that for each particle, the bond and the dual-bond are independent from each other; the same applies to the horizon and dual-horizon. It means the bond and dual-bond can break independently in the fracture models. For discretisations with varying horizons, often one particle is within the horizons of other particles but not vice versa. In this case, a single horizon is not sufficient to define the interactions between particles. The concept of two horizons proposed here can solve this issue with a simple and direct physical meaning. It naturally takes into account the interactions between particles of varying horizon sizes. For models with constant horizons, horizon and dual-horizon will degenerate to the horizon in original peridynamics. \vspace{3mm}

\begin{figure}[htp]
	\centering
		\includegraphics[width=9cm]{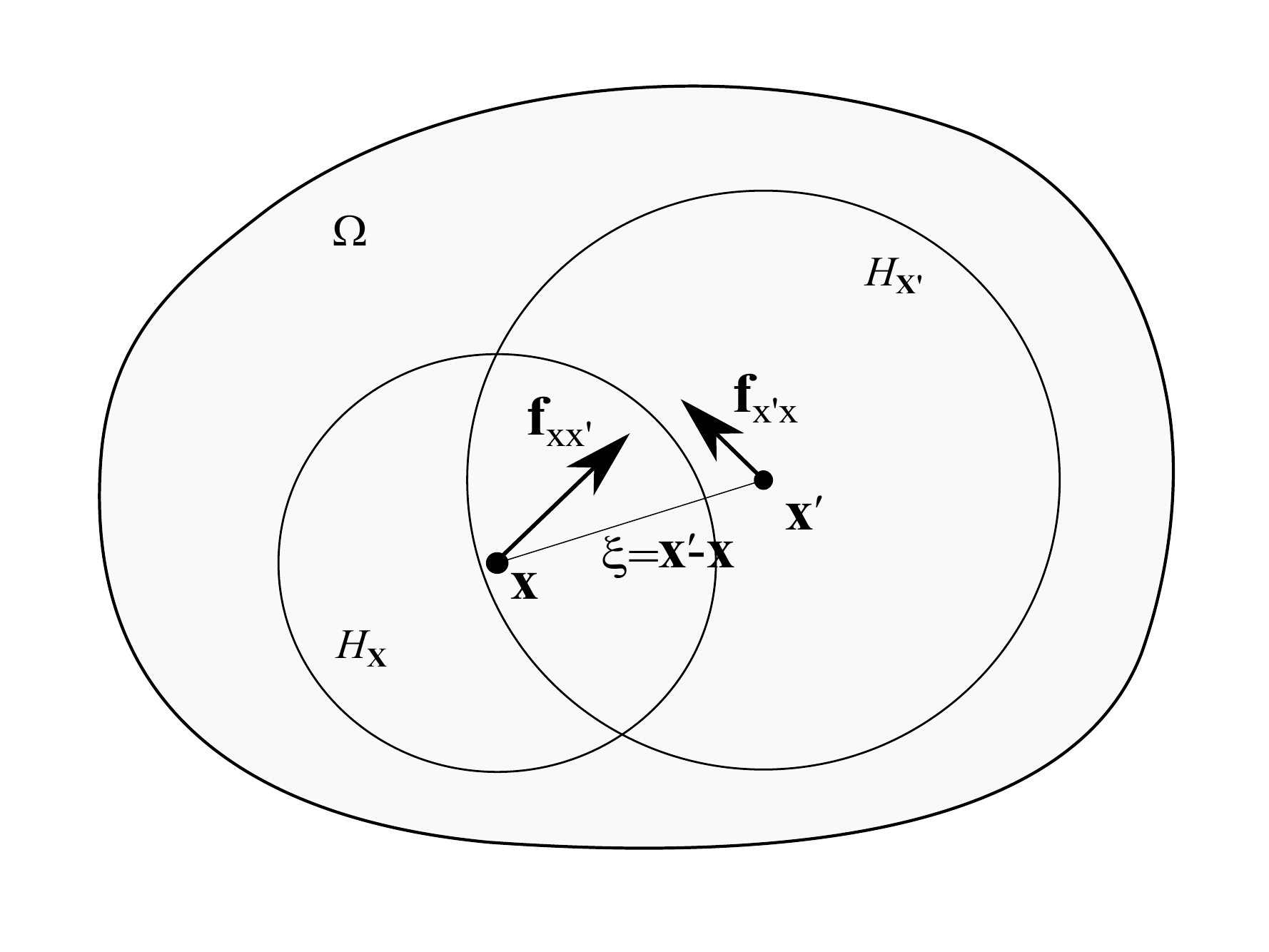}
	\caption{Double summation of force}
	\label{fig:ProofFxx}
\end{figure}

\noindent \textbf{Reaction force by horizon and dual-horizon}\\
In our dual-horizon peridynamic formulation, the horizons are differentiated between how a particle receives and exerts forces with other particles. Under this new concept, computing the force $\bff_{\bx \bx'}$ between a pair of particles, denoted as particles $\bx$ and $\bx'$, is determined by whether $\bx$ is within the horizon of $\bx'$, and whether $\bx'$ is inside the dual-horizon of $\bx$, and vice versa for $\bff_{\bx' \bx}$. It can be easily seen that the force density vector has the following property,
\begin{align}
\text{if} \quad \bx \in H_{\bx'} \,\text{or} \, \bx' \in H_{\bx}^{'}, \bff_{\bx \bx'} \ne 0 \ , \notag\\
\text{else} \quad \bff_{\bx \bx'} = 0  \ .
\end{align}
For any bond between two particles $\bx $ and $\bx' $ belonging to a domain denoted as $\bB$ in Fig.\ref{fig:ProofFxx}, the direct force $\bff_{\bx \bx'}$ acting on $\bx $ due to $\bx' $ can be computed by two approaches as follows. \\
\textbf{Approach 1} computes the force in terms of $\bx $,
\begin{align}
\bff_{\bx \bx'} = \begin{cases} \ne 0 &\mbox{if } \bx' \in H_{\bx}' \\
0 & \mbox{if } \bx' \notin H_{\bx}' \end{cases} \notag \ ,
\end{align}
and \textbf{Approach 2} is formulated with respect to $\bx'$,
\begin{align}
\bff_{\bx \bx'} = \begin{cases} \ne 0 &\mbox{if } \bx \in H_{\bx'} \\
0 & \mbox{if} \bx \notin H_{\bx'} \end{cases} \notag \ .
\end{align}

For any domain under consideration e.g. the shaded area in Fig.\ref{fig:ProofFxx}, the computation of the direct forces that take place between particles (no reactive force considered yet) shall undertake all forces from any particle that belongs to $\bB$. There are two approaches to achieve that. The first approach is by summing all the forces the particles undertake in the dual-horizon,
\begin{align}
\label{eq:force1}
\sum_{\bx \in \bB} \left(\sum_{\bx' \in H_{\bx}'} \bff_{\bx \bx'}\, \triangle V_{\bx'}\right) \triangle V_{\bx} .
\end{align}
And the second is to add up all the forces of each particle that it applies to other particles,
\begin{align}
\label{eq:force2}
\sum_{\bx' \in \bB} \left(\sum_{\bx \in H_{\bx'}} \bff_{\bx \bx'}\, \triangle V_{\bx}\right) \triangle V_{\bx'} \ .
\end{align}
Since the total force for any $\bB$ is independent of the approach chosen to compute it, Eq. (\ref{eq:force1}) and (\ref{eq:force2}) shall be equal
\begin{align}
\label{eq:equalforce}
\sum_{\bx \in \bB} \sum_{\bx' \in H_{\bx}'} \bff_{\bx \bx'}\, \triangle V_{\bx'} \triangle V_{\bx}=\sum_{\bx' \in \bB} \sum_{\bx \in H_{\bx'}} \bff_{\bx \bx'}\, \triangle V_{\bx} \triangle V_{\bx'}  \ .
\end{align}
Eq. (\ref{eq:equalforce}) indicates that changing the summation order from $\bx \to \bx'$ shall be done together with changing from $H_{\bx'} \to H_{\bx}'$ so as to keep all the variables consistent. When the discretisation is sufficiently fine, the summation shall approximate to the integral and Eq. (\ref{eq:equalforce}) becomes
\begin{align}
\int_{\bx \in \bB} \int_{\bx' \in H_{\bx}'} \bff_{\bx \bx'}\, \ud V_{\bx'} \ud V_{\bx}= \int_{\bx' \in \bB} \int_{\bx \in H_{\bx'}} \bff_{\bx \bx'}\, \ud V_{\bx} \ud V_{\bx'}\ .\label{eq:doubleInt}
\end{align}

\subsection{Motion equation for peridynamics with horizon-variable}\label{sec:MotionEquation}
In the following derivation $\bx'-\bx$ will be denoted as $\bxi$, and $\bu(\bx',t)-\bu(\bx,t)$ as $\boldeta$, hence $\bxi+\boldeta$ represents the current relative position vector between the particles. The internal forces that are exerted at each particle from the other particles should include two parts, namely the forces from the horizon and the forces from the dual-horizon. The other forces applied to a particle include the body force and the inertia force. Let $\triangle V_{\bx}$ denote the volume associates to $\bx$. The body force for particle $\bx $ can be expressed as $\mathbf b(\bx, t) \triangle V_{\bx}$, where $ \mathbf b(\bx,t)$ is the body force density. The inertia is denoted by $\rho \mathbf{\ddot{u}}(\bx,t) \triangle V_{\bx}$, where $\rho$ is the density associated to $\bx$. At any time $t$, for $\bx'$ in dual-horizon of $\bx$, the force vector of $\tilde{\bff}_{\bx \bx'}$ is defined as
\begin{align}
\label{eq:pactivef}
\tilde{\bff}_{\bx \bx'}:= \bff_{\bx \bx'}(\boldeta,\bxi) \cdot \triangle V_{\bx} \cdot \triangle V_{\bx'} \ ,
\end{align}
where $\bff_{\bx \bx'}(\boldeta,\bxi)$ is the force density in the traditional peridynamics with unit of force per volume squared; $\tilde{\bff}_{\bx \bx'}$ is the force vector acting on particle $\bx$ due to the attraction or repulsion from $\bx'$. The forces from the dual-horizon are active forces as they are applied to $\bx$. The total force applied to $\bx$ from its dual-horizon, $H_{\bx}'$, can be computed by,
\begin{align}
\sum_{\bx' \in H_{\bx}'} \tilde{\bff}_{\bx \bx'}
=\sum_{\bx' \in H_{\bx}'}\bff_{\bx \bx'}(\boldeta,\bxi) \cdot \triangle V_{\bx} \cdot \triangle V_{\bx'} \ .
\label{eq:activef}
\end{align}
For any $\bx'$ inside the horizon of $\bx$, the force $\tilde{\bff}_{\bx' \bx}$ acting on $\bx'$ due to $\bx $ is defined as
\begin{align}
\tilde{\bff}_{\bx' \bx}:= \bff_{\bx' \bx}(-\boldeta,-\bxi) \cdot \triangle V_{\bx'} \cdot \triangle V_{\bx}
\label{eq:ppassivef}
\end{align}
where $\bff_{\bx' \bx}(-\boldeta,-\bxi) $ is the force density per volume squared; $\tilde{\bff}_{\bx' \bx}$ is the force vector acting on particle $\bx'$ due to the attraction or repulsion from $\bx$ ( $\bx'$ inside horizon of $\bx$). The total force $\bx$ exerted on other particles from horizon $H_{\bx}$ is the summation of $\tilde{\bff}_{\bx' \bx}$
\begin{align}
\sum_{\bx' \in H_{\bx}} \tilde{\bff}_{\bx' \bx}
= \sum_{\bx' \in H_{\bx}}\bff_{\bx' \bx}(-\boldeta,-\bxi) \cdot \triangle V_{\bx}\cdot \triangle V_{\bx'}
\end{align}
Based on Newton's third law, the total force $\bx$ undertaken in $H_{\bx}$ takes opposite direction
\begin{align}
-\sum_{\bx' \in H_{\bx}} \tilde{\bff}_{\bx' \bx}
=-\sum_{\bx' \in H_{\bx}}\bff_{\bx' \bx}(-\boldeta,-\bxi) \cdot \triangle V_{\bx}\cdot \triangle V_{\bx'}
\label{eq:npassivef}
\end{align}
By summing over all forces on particle $\bx$, including inertia force, body force, active force in Eq. (\ref{eq:activef}) and passive force in Eq. (\ref{eq:npassivef}), we obtain the equations of motion
\begin{align}
\rho \mathbf{\ddot{u}}(\bx,t) \triangle V_{\bx} =
\sum_{\bx' \in H_{\bx}'}\tilde{\bff}_{\bx \bx'}+ \left(-\sum_{\bx' \in H_{\bx}}\tilde{\bff}_{\bx' \bx}\right) + \mathbf b(\bx,t) \triangle V_{\bx}
\label{eq:discrete1}  \ .
\end{align}
Substituting Eqs. (\ref{eq:activef})  and (\ref{eq:npassivef}) into Eq. (\ref{eq:discrete1}) leads to
\begin{align}
\rho \mathbf{\ddot{u}}(\bx,t) \triangle V_{\bx}
=\sum_{\bx' \in H_{\bx}'} \bff_{\bx \bx'}(\boldeta,\bxi) \triangle V_{\bx'}  \triangle V_{\bx}-\sum_{\bx' \in H_{\bx}}\bff_{\bx' \bx}(-\boldeta,-\bxi) \triangle V_{\bx'}  \triangle V_{\bx}+\mathbf b(\bx,t) \triangle V_{\bx}\label{eq:discrete2} \ .
\end{align}
As the volume $\triangle V_{\bx} $ associated to particle $\bx$ is independent of the summation, we can eliminate $\triangle V_{\bx} $ in Eq. (\ref{eq:discrete2}), yielding the governing equation based on $\bx$:
\begin{align}
\rho \mathbf{\ddot{u}}(\bx,t)
=\sum_{\bx' \in H_{\bx}'} \bff_{\bx \bx'}(\boldeta,\bxi) \triangle V_{\bx'}-\sum_{\bx' \in H_{\bx}}\bff_{\bx' \bx}(-\boldeta,-\bxi) \triangle V_{\bx'}+\mathbf b(\bx,t) \ .
\label{eq:discreteMotion}
\end{align}
When the discretisation is sufficiently fine, the summation is approximating to the integration of the force on the dual-horizon and horizon,
\begin{align}
\lim_{\triangle V_{\bx'} \to 0}\sum_{\bx' \in H_{\bx}'} \bff_{\bx \bx'}(\boldeta,\bxi)  \triangle V_{\bx'}
=\int_{\bx' \in H_{\bx}'} \bff_{\bx \bx'}(\boldeta,\bxi) \, \ud V_{\bx'}
\end{align}
and
\begin{align}
\lim_{\triangle V_{\bx'} \to 0}\sum_{\bx' \in H_{\bx}}\bff_{\bx' \bx}(-\boldeta,-\bxi) \triangle V_{\bx'}
=\int_{\bx' \in H_{\bx}}\bff_{\bx' \bx}(-\boldeta,-\bxi) \, \ud V_{\bx'} \ .
\end{align}
Thus the integration form of the equation of motion in peridynamics with dual horizon is given as
\begin{align}
\rho \mathbf{\ddot{u}}(\bx,t)
=\int_{\bx' \in H_{\bx}'} \bff_{\bx \bx'}(\boldeta,\bxi) \, \ud V_{\bx'} -\int_{\bx' \in H_{\bx}}\bff_{\bx' \bx}(-\boldeta,-\bxi) \, \ud V_{\bx'} +\mathbf b(\bx,t) \ .
\label{eq:intMotion}
\end{align}
Eq. (\ref{eq:intMotion}) is similar to Eq. (\ref{eq:oldPD}) of the original peridynamics theory. When the horizons are set constant, i.e. both horizon and the dual-horizon are equal, the integrations in Eq. (\ref{eq:intMotion}) degenerate to the original peridynamics theory. It means the traditional peridynamics can be viewed as a special case of the present dual horizon peridynamics.

About the implementation of the present peridynamic formulation, for any particle $\bx$, the force density $\bff_{\bx \bx'}(\boldeta,\bxi)$ in $H_{\bx}'$ can be determined when calculating the force in $H_{\bx'}$ for particle $\bx'$. Therefore, it is not necessary to know exactly the dual-horizon geometry and the formulation can be implemented with minor modification of any peridynamic codes.

\subsection{Proof of basic physical principles}\label{sec:Proof}
\subsubsection{Balance of linear momentum }\label{subsec:LinearMomentum}
The internal forces shall satisfy the balance of linear momentum for any bounded body $\bB $ given by
\begin{align}
&\int_\bB (\rho \mathbf{\ddot{u}}(\bx,t)-\mathbf b(\bx,t)) \ud V_{\bx}\notag\\
&=\int_\bB \int_{\bx' \in H_{\bx}'} \bff_{\bx \bx'}(\boldeta,\bxi) \, \ud V_{\bx'} \ud V_{\bx}-\int_\bB \int_{\bx' \in H_{\bx}}\bff_{\bx' \bx}(-\boldeta,-\bxi) \, \ud V_{\bx'} \ud V_{\bx} \notag\\
&=\mathbf 0 \label{eq:lineareq} \ .
\end{align}
For simplicities, let $\bff_{\bx \bx'} $ represent $ \bff_{\bx \bx'}(\boldeta,\bxi) $, and $\bff_{\bx' \bx} $ the $\bff_{\bx' \bx}(-\boldeta,-\bxi) $.

\textbf{Proof}:
For convenience, we start with the discrete form of the Eq. (\ref{eq:discreteMotion}) as
\begin{align}
\label{eq:prooflinear}
\sum_{\bx \in \bB} (\rho \mathbf{\ddot{u}}(\bx,t)-\mathbf b(\bx,t)) \triangle V_{\bx} &=\sum_{\bx \in \bB} \sum_{\bx' \in H_{\bx}'} \bff_{\bx \bx'}\, \triangle V_{\bx'} \triangle V_{\bx}
-\sum_{\bx \in \bB} \sum_{\bx' \in H_{\bx}} \bff_{\bx' \bx}\, \triangle V_{\bx'} \triangle V_{\bx}
\end{align}

The first term on RHS of Eq. (\ref{eq:prooflinear}) can be substituted with Eq. (\ref{eq:equalforce}) by changing the summation domain and therefore we will get,
\begin{align}
\sum_{\bx \in \bB} (\rho \mathbf{\ddot{u}}(\bx,t)-\mathbf b(\bx,t)) \triangle V_{\bx} =\sum_{\bx' \in \bB} \sum_{\bx \in H_{\bx'}} \bff_{\bx \bx'}\, \triangle V_{\bx'} \triangle V_{\bx}
-\sum_{\bx \in \bB} \sum_{\bx' \in H_{\bx}} \bff_{\bx' \bx}\, \triangle V_{\bx'} \triangle V_{\bx} \ ,
\end{align}
The above transformation is based on the idea that $\bff_{\bx \bx'}$ acting on $\bx $ is computed by using dual-horizon ($H_{\bx}'$) of $\bx $. And $\bff_{\bx \bx'}$ is calculated by using  horizon ($H_{\bx'}$) of $\bx'$. Note that in both calculations, $\bff_{\bx \bx'}$ remains unchanged and only the definition domain changes. Since the dummy variables can be relabeled for $\bx \leftrightarrow \bx' $ in the first term on RHS, thus the first term has the same expression as the second term, i.e.
\begin{align}
\sum_{\bx \in \bB} (\rho \mathbf{\ddot{u}}(\bx,t)-\mathbf b(\bx,t)) \triangle V_{\bx}=\sum_{\bx \in \bB} \sum_{\bx' \in H_{\bx}} \bff_{\bx' \bx}\, \triangle V_{\bx} \triangle V_{\bx'}
-\sum_{\bx \in \bB} \sum_{\bx' \in H_{\bx}} \bff_{\bx' \bx}\, \triangle V_{\bx'} \triangle V_{\bx}\notag = 0
\end{align}

As the forces are differentiated here between active and passive force, i.e. an active force from $\bx$ corresponds to a passive force of $\bx'$, the forces is always pairwise but with opposite direction and of the same magnitude. Hence, it is natural that the balance of linear momentum is satisfied.
\subsubsection{Balance of angular momentum }\label{subsec:AngularMomentum}
Let $\by $ be the deformation vector state field defined by
\begin{align}
\by[\bx,t]\langle \bxi \rangle=\by(\bx',t)- \by(\bx,t) \qquad \forall \bx \in \bB, \bxi=\bx'-\bx, \bx' \in H_\bx, t \ge 0 \,
\end{align}
\begin{align}
\bu(\bx,t)=\by(\bx,t)- \bx  \ ,
\end{align}
where $\by(\bx,t)$ refers to the current configuration coordinate for $\bx$ in material configuration, $\bu(\bx,t)$ is the displacement for $\bx $, $H_\bx $ is the horizon. Thus $\by\langle \bx'-\bx \rangle $ is the image of the bond $ \bx'-\bx $ under the deformation.

To  satisfy the balance of angular momentum for any bounded body $\bB $,it is  required that:
\begin{align}
\int_\bB &\by \times (\rho \mathbf{\ddot{u}}(\bx,t)-\mathbf b(\bx,t)) \ud V_{\bx} \notag\\
&=\int_\bB \by \times \left(\int_{\bx' \in H_{\bx}'} \bff_{\bx \bx'}(\boldeta,\bxi) \, \ud V_{\bx'} -\int_{\bx' \in H_{\bx}}\bff_{\bx' \bx}(-\boldeta,-\bxi) \, \ud V_{\bx'}\right) \ud V_{\bx}\notag\\
&=\int_\bB \int_{\bx' \in H_{\bx}'} \by \times \bff_{\bx \bx'}(\boldeta,\bxi) \, \ud V_{\bx'} \ud V_{\bx} -\int_\bB \int_{\bx' \in H_{\bx}}\by \times\bff_{\bx' \bx}(-\boldeta,-\bxi) \, \ud V_{\bx'} \ud V_{\bx}\notag\\
&=\mathbf 0
\label{eq:Angreq}
\end{align}
For simplicities, let $\bff_{\bx \bx'} $ represent $ \bff_{\bx \bx'}(\boldeta,\bxi) $, and $\bff_{\bx' \bx} $ present $\bff_{\bx' \bx}(-\boldeta,-\bxi) $.

\noindent \textbf{Proposition}: In the dual-horizon peridynamics, suppose a constitutive model of the form
\begin{align}
\bff =\hat{\bff}(\by,\Lambda)
\end{align}
where $ \hat{\bff }: \mathcal V\to \mathcal V\ $ is bounded and Riemann-integrable on $H $ and $\mathcal V $ is the vector state; $\Lambda $ denotes all variables other than the current deformation vector state that $\bf $ may depend on for some particular material.

\noindent  If
\begin{align}
\int_{\bx' \in H_{\bx}} \by\langle \bx'-\bx \rangle \times \bff_{\bx' \bx}\, \ud V_{\bx'}=0 \qquad \forall \by \in \mathcal V \ ,
\end{align}
or in discrete form
\begin{align}
\sum_{\bx' \in H_{\bx}} \by\langle \bx'-\bx \rangle \times \bff_{\bx' \bx}\, \triangle V_{\bx'}=0 \qquad \forall \by \in \mathcal V \ ,
\end{align}
where $\bff_{\bx' \bx}= \bff_{\bx' \bx}(\bu(\bx',t)-\bu(\bx,t),\bx'-\bx) $ , the force vector density acting on $\bx' $, and $H_{\bx} $ is the horizon of $\bx$. Then, the balance of angular momentum, Eq. (\ref{eq:Angreq}) holds for any deformation of $\bB $ for any given constitutive model.

\textbf{Proof}:
As the summation over the domain $\bB $ is equivalent to the integrand of the same expression over that domain, we use the discrete form as
\begin{align}
\label{eq:proofangular}
& \sum_{\bx \in \bB} \by \times (\rho \mathbf{\ddot{u}}(\bx,t)-\mathbf b(\bx,t)) \triangle V_{\bx}\notag\\
&=\sum_{\bx \in \bB} (\bx+\bu) \times (\rho \mathbf{\ddot{u}}(\bx,t)-\mathbf b(\bx,t)) \triangle V_{\bx} \notag\\
&=\sum_{\bx \in \bB} \sum_{\bx' \in H_{\bx}'}(\bx+\bu) \times \bff_{\bx \bx'}\, \triangle V_{\bx'} \triangle V_{\bx}
-\sum_{\bx \in \bB} \sum_{\bx' \in H_{\bx}} (\bx+\bu) \times \bff_{\bx' \bx}\, \triangle V_{\bx'} \triangle V_{\bx}
\end{align}
The first term on the RHS of Eq. (\ref{eq:proofangular}) can be rewritten according to Eq. (\ref{eq:equalforce}) by changing the summation domain and thus it becomes
\begin{align}
&\sum_{\bx \in \bB} \by \times (\rho \mathbf{\ddot{u}}(\bx,t)-\mathbf b(\bx,t)) \triangle V_{\bx}\notag\\
&=\sum_{\bx' \in \bB} \sum_{\bx \in H_{\bx'}}(\bx+\bu) \times \bff_{\bx \bx'}\, \triangle V_{\bx'} \triangle V_{\bx}
-\sum_{\bx \in \bB} \sum_{\bx' \in H_{\bx}} (\bx+\bu) \times \bff_{\bx' \bx}\, \triangle V_{\bx'} \triangle V_{\bx} \ .
\end{align}
Swapping the dummy variables for the first term on RHS between $\bx \leftrightarrow \bx' $ in the double summation, then the first term takes the same summation to the second term on RHS
\begin{align}
&\sum_{\bx \in \bB} \by \times (\rho \mathbf{\ddot{u}}(\bx,t)-\mathbf b(\bx,t)) \triangle V_{\bx}\notag\\
&=\sum_{\bx \in \bB} \sum_{\bx' \in H_{\bx}} (\bx'+\bu') \times \bff_{\bx' \bx}\, \triangle V_{\bx} \triangle V_{\bx'}
-\sum_{\bx \in \bB} \sum_{\bx' \in H_{\bx}} (\bx+\bu) \times \bff_{\bx' \bx}\, \triangle V_{\bx'} \triangle V_{\bx}\notag\\
&=\sum_{\bx \in \bB} (\sum_{\bx' \in H_{\bx}} ((\bx'+\bu')-(\bx+\bu)) \times \bff_{\bx' \bx}\, \triangle V_{\bx'}) \triangle V_{\bx}\notag\\
&=\sum_{\bx \in \bB} (\sum_{\bx' \in H_{\bx}} (\by'-\by) \times \bff_{\bx' \bx}\, \triangle V_{\bx'}) \triangle V_{\bx}\notag\\
&=\sum_{\bx \in \bB} (\sum_{\bx' \in H_{\bx}} \by\langle \bx'-\bx \rangle \times \bff_{\bx' \bx}\, \triangle V_{\bx'}) \triangle V_{\bx}=\mathbf 0  \, 
\end{align}
or in an integration form for sufficiently fine discretisation as
\begin{align}
\int_{\bx \in \bB} \by \times (\rho \mathbf{\ddot{u}}(\bx,t)-\mathbf b(\bx,t)) \ud V_{\bx}=\mathbf 0 \ .
\end{align}
Therefore, the angular momentum over the entire analysis domain is satisfied. $\square $

\begin{figure}[htp]
	\centering
		\includegraphics[width=9cm]{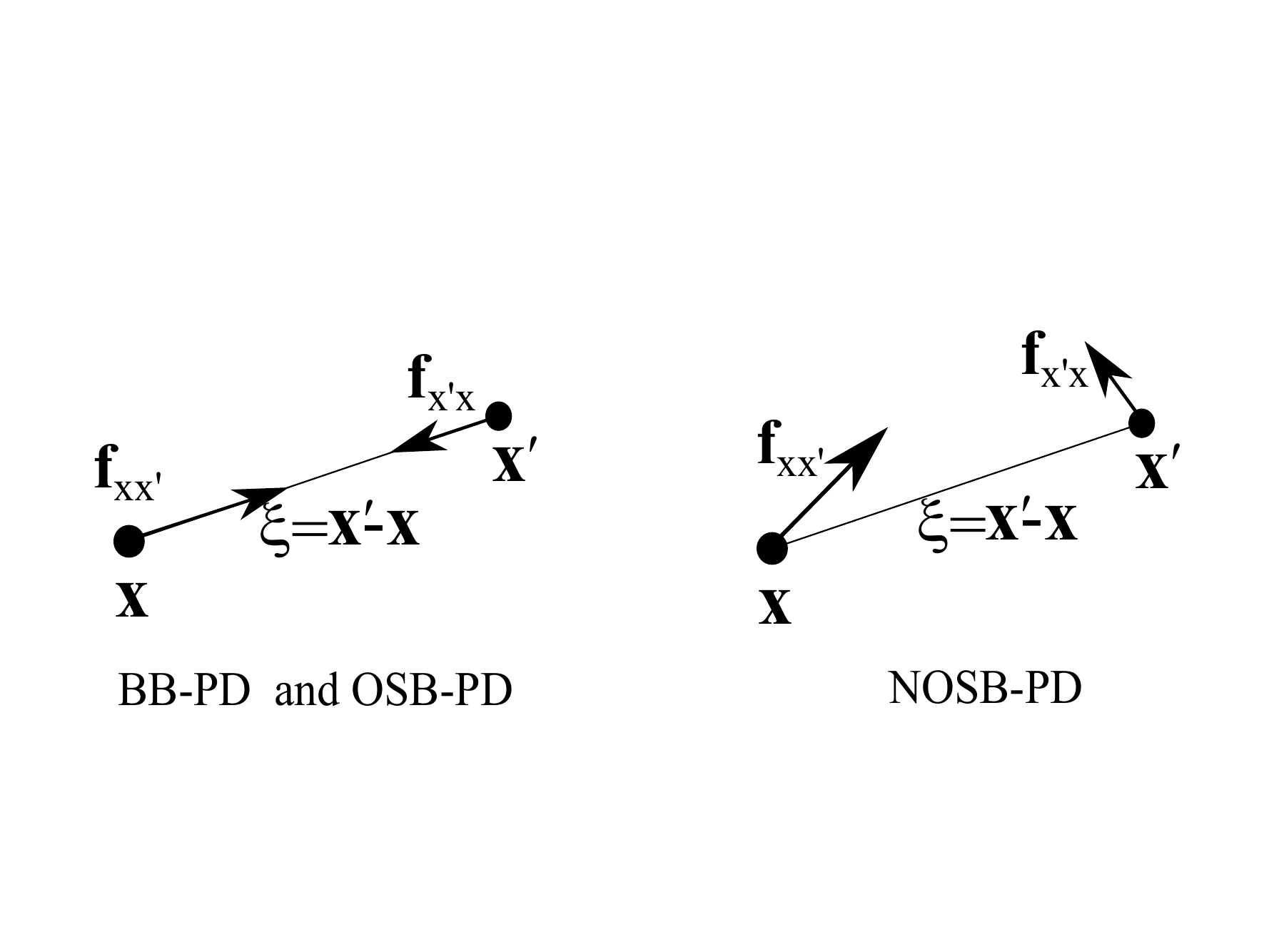}
	\caption{force vector of BB-PD,OSB-PD and NOSB-PD}
	\label{fig:3PD}
\end{figure}

It can be seen that the conservation of angular somehow depends \textbf{only} on the horizon; the dual-horizon is not involved. The latter is only needed in the Eq. (\ref{eq:discreteMotion}) and Eq. (\ref{eq:intMotion}). The bond-based, ordinary based and non-ordinary based peridynamics all satisfy the angular momentum in the original horizon concept (see proof in \cite{Silling2007}). This conclusion can be also illustrated for the BB-PD and OS-PD since the internal forces $\bff_{\bx \bx'} $ and $\bff_{\bx' \bx} $ are parallel to the bond vector in the current configuration, see Fig.\ref{fig:3PD}.

\section{Dual-horizon peridynamics}
The dual-horizon formulation will now be applied to all existing peridynamics, namely BB-PD, OSB-PD and NOSB-PD. The calibration of constitutive parameters with respect to the continuum model and some issues concerning the implementations will be discussed for 2D and 3D problems.

\subsection{Dual-horizon bond based peridynamics}\label{subsec:BBPD}
In the bond-based peridynamics theory \cite{Gerstle2007,Silling2005}, the pair force function is expressed as
\begin{align}
\bff(\boldeta,\bxi) =\frac{ \partial w(\boldeta,\bxi)}{\partial \boldeta}\qquad \forall \boldeta,\bxi
\end{align}
For a micro-elastic homogenous and isotropic material, $\bff(\boldeta,\bxi)$ can be specified as
\begin{align}
\bff(\boldeta,\bxi) =c s \cdot \frac{ \boldeta +\bxi}{\|{ \boldeta+\bxi}\|} \ ,
\end{align}
where $c$ is the micro-modulus, and $s$ is the bond stretch calculated by
\begin{align}
s=\frac{\|\boldeta+\bxi\|-\|\bxi\|}{\|\bxi\|} \ .
\end{align}
The energy per unit volume in the body at a given time $t$ is given by \cite{Silling2005}
\begin{align}
W=\frac{1}{2}\int_{H_\bx} w(\boldeta,\bxi) \ud V_{\bxi} \ .
\end{align}
For the BB-PD theory, by enforcing the strain energy density being equal to the strain energy density in the classical theory of elasticity \cite{Gerstle2007, Silling2005}, and by setting the influence function $\omega(\|{\bxi}\|)=1$, we obtain
\begin{align}
\label{eq:cdelta}
C(\delta)&=\frac{3E}{\pi \delta^3 (1-\nu)} \quad &\mathrm{plane\,stress} \notag\\
C(\delta)&=\frac{3E}{\pi \delta^3 (1+\nu) (1-2\nu)} \quad &\mathrm{plane\,strain} \notag\\
C(\delta)&=\frac{3E}{\pi \delta^4 (1-2\nu)} \quad &\mathrm{3D}  \ .
\end{align}
Note that the value of $C(\delta)$ takes half of the micro-modulus $c$ used in the horizon-constant BB-PD since the bond energy for varying horizon is determined by both horizon and dual-horizon. It can be easily seen that when the horizon takes the same value as the dual-horizon, the bond energy between two particles is reduced to that of the original BB-PD.

Let $w_0(\bxi)=C(\delta) s_0^2(\delta) \xi/2 $ denote the work required to break a single bond, where $s_0(\delta) $ is the critical bond stretch. By breaking half of all the bonds connected to a given particle along the fracture surface and equalizing the breaking bonds energy with the energy release rate $G_0 $ \cite{Dipasquale2014, Gerstle2007}, we can get the expression between the energy release rate $G_0 $ and the critical bond stretch $s_0(\delta) $:	
\begin{align}
\label{eq:criticalStr}
s_0(\delta)&=\sqrt{\frac{4 \pi G_0}{9 E \delta}} \quad &\mathrm{plane\,stress} \notag\\
s_0(\delta)&=\sqrt{\frac{5 \pi G_0}{12 E \delta}} \quad &\mathrm{plane\,strain} \notag\\
s_0(\delta)&=\sqrt{\frac{5 G_0}{6 E \delta}} \quad &\mathrm{3D} \ .
\end{align}
Both the micro-modulus $C(\delta) $ and the critical stretch $s_0(\delta) $ are derived from the local continuum mechanics theory, and they depend on the horizon radii for variable horizons.

In the implementation of the bond-based peridynamics, fracture is introduced by removing particles from the neighbour list once the bond stretch exceeds the critical bond stretch $s_0 $. In order to specify whether a bond is broken or not, a history-dependent scalar valued function $\mu$ is introduced \cite{Silling2005},
\begin{align}
\mu(t,\bxi)=
\begin{cases}
1 \,\qquad\mathrm{if}\, s(t',\bxi)<s_0 \,\,\text{for all }\, 0 \le t' \le t,\\
0 \,\qquad\mathrm{otherwise.}
\end{cases}
\end{align}
The local damage at $\bx$ is defined as
\begin{align}
\label{eq:damage}
\phi(\bx, t)=1-\frac{\int_{H_{\bx}} \mu(\bx, t,\bxi) \ud V_{\xi}}{\int_{H_{\bx}} \ud V_{\xi}} \ .
\end{align}
The damage formulation of Eq. (\ref{eq:damage}) is also applicable to OSB-PD. For any particle with dual-horizon ($H_{\bx}'$) and horizon ($H_{\bx}$), the active force $\bff_{\bx \bx'} $ and the passive force $\bff_{\bx' \bx}$ in Eq. (\ref{eq:discreteMotion}) or (\ref{eq:intMotion}) are computed by the following expressions, respectively
\begin{align}
\bff_{\bx \bx'}=C(\delta_{\bx'}) \cdot s_{\bx \bx'} \cdot \frac{ \boldeta +\bxi}{\|{ \boldeta+\bxi}\|},\qquad \forall \bx' \in H_{\bx}'
\end{align}
\begin{align}
\bff_{\bx' \bx}=C(\delta_{\bx}) \cdot s_{\bx \bx'} \cdot \frac{-( \boldeta +\bxi)}{\|{ \boldeta+\bxi}\|},\qquad \forall \bx' \in H_{\bx} \ ,
\end{align}
where $C(\delta_{\bx'})$ and $C(\delta_{\bx})$ are the micro-modulus based on $\delta_{\bx'}$ and $\delta_{\bx} $ computed from Eq. (\ref{eq:cdelta}) respectively, and $s_{\bx \bx'}$ is the stretch between particles $\bx$ and $\bx'$.

\subsection{Dual-horizon ordinary state based peridynamics}\label{subsec:UOSBPD}
The concept of ``state" for peridynamics was firstly introduced in \cite{Silling2007}. A state of order $m$ is a function
\begin{equation}
\underline{\mathbf A}: H \to \mathbf T_m,\bxi \mapsto \underline{\mathbf A}\langle \bxi \rangle.
\end{equation}
where $H$ is the horizon domain, $\mathbf T_m$ denotes the set of all tensors of order $m$, $\langle \square \rangle$ indicate the vector to which a state operates. In this paper, the scalar state and vector state are use if not otherwise specified.

The 2D OSB-PD formulation of bond force can be derived by following the similar procedure as that was described in 3D.
Both 2D and 3D OSB-PD can be written in a unified expression taking into account the dimension number as
\begin{align}
\label{eq:unifiedOSBPD}
\underline{t}\langle \bxi \rangle =\frac{n K \theta}{m} \omega\langle\bxi\rangle \cdot \xi+\frac{n(n+2)G}{m} \omega\langle\bxi\rangle\underline{e}^d\langle \bxi \rangle \ ,
\end{align}
where $n\in \{2,3\} $ is the dimensional number, $K $ is the bulk modulus, $G $ the shear modulus, $\displaystyle \xi=\|\bxi\|$, $\underline{e}^d \langle \bxi \rangle =\underline{e}\langle \bxi \rangle-\displaystyle{\frac{\theta \xi}{n}} $, $m=\int_H \omega\langle\bxi\rangle \,\bxi \cdot \bxi\, \ud V_{\bxi}$, $\displaystyle{\theta=\frac{n}{m}} \int_H{\omega\langle\bxi\rangle \, \xi \cdot \underline{e}\langle \bxi \rangle \ud V_{\bxi}}$, $\underline{e}\langle \bxi \rangle= \|\bxi+\boldeta\|-\|\bxi\|$ and $\ud V_{\xi}$ are the area in 2D, and volume in 3D, respectively. For any particle $\bx$, $\bff_{\bx \bx'}$ and $\bff_{\bx' \bx}$ in Eqs. (\ref{eq:discreteMotion}) and (\ref{eq:intMotion}) can be calculated by
\begin{align}
\bff_{\bx \bx'}=\underline{t} \langle \bxi \rangle \cdot \frac{\boldeta +\bxi}{\|{\boldeta+\bxi}\|},\qquad \forall \bx' \in H_{\bx}' \ ,
\end{align}
and
\begin{align}
\bff_{\bx' \bx}=\underline{t}' \langle -\bxi \rangle \cdot \frac{ -\boldeta -\bxi}{\|{ \boldeta+\bxi}\|},\qquad \forall \bx' \in H_{\bx} \ .
\end{align}
Here $\bxi =\bx'-\bx $, and $\bx$ and $\bx' $ are the coordinate vectors for $\bx$ and particle $\bx'$ in the material configuration, respectively.  

\subsection{Dual-horizon non-ordinary state based peridynamics}\label{subsec:NOSBPD}
NOSB-PD uses the deformation gradient from classic continuum mechanics. It offers the possibility to consistently include existing constitutive models. With the use of the shape tensor proposed in the NOSB-PD, the fundamental tensors in continuum mechanics can be easily introduced in  peridynamics, including the deformation gradient tensor or velocity gradient \cite{Silling2010, Warren2008}. The shape tensor for particle $\bx$ is defined as
\begin{align}
\label{eq:shapetensor}
\mathbf K_\bx&=\int_{\bx' \in H_{\bx}} \omega(\bxi) \cdot \bxi \otimes \bxi \ud V_{\bx'} \,
\end{align}
where $\bxi=\bx'-\bx $ is the bond in the reference configuration, $H_{\bx}$ is the \textit{horizon} for particle $\bx$ and  $\omega(\bxi) $ is the influence function. The deformation tensor for particle $\bx$ is given by
\begin{align}
\mathbf F_{\bx}=\frac{\partial{\by}}{\partial{\bx}}=\int_{\bx' \in H_{\bx}} \omega(\bxi)\by
\langle \bxi \rangle \otimes \bxi \; \ud V_{\bx'}\,\cdot \mathbf K_{\bx}^{-1} \ ,
\end{align}
where $\by:=\by(\bx,t) $ are the spatial coordinates, $\bx$ are the material coordinates, and $\bxi=\bx'-\bx$. The spatial velocity gradient $\mathbf L_{\bx} $ for particle $\bx $ is defined in the current configuration using the chain rule,
\begin{align}
\label{eq:localvelocityGradient}
\mathbf L_{\bx}:=\frac{\partial{\mathbf v}}{\partial{\by}}=\frac{\partial{\mathbf v}}{\partial{\bx}}\cdot \frac{\partial{\bx}}{\partial{\by}}=\dot{\mathbf F}_\bx \mathbf F_{\bx}^{-1} \ ,
\end{align}
where $\mathbf v:=\mathbf v(\bx,t)=\displaystyle{\frac{\partial \by}{\partial t}}(\bx,t)$ is the velocity vector and $\dot{\mathbf F}_\bx $ is the rate of deformation gradient. The non-local form of the velocity gradient can be written as
\begin{align}
\label{eq:nonlocalvelocityGradient}
\mathbf L_{\bx}:&=\int_{\bx' \in H_{\bx}} \omega(\bxi )\,{\mathbf v}
\langle \bxi \rangle \otimes \bxi \;\ud V_{\bx'}\,\cdot \mathbf K_{\bx}^{-1} \cdot
 \left( \int_{\bx' \in H_{\bx}} \omega(\bxi)\,\by
\langle \bxi \rangle \otimes \bxi \;\ud V_{\bx'}\,\cdot \mathbf K_{\bx}^{-1} \right )^{-1}\notag\\
&= \int_{\bx' \in H_{\bx}} \omega(\bxi )\,\mathbf v
\langle \bxi \rangle \otimes \bxi \;\ud V_{\bx'}\,\cdot \mathbf K_{\bx}^{-1}\cdot \mathbf K_{\bx}
\left( \int_{\bx' \in H_{\bx}} \omega(\bxi)\,\by
\langle \bxi \rangle \otimes \bxi \;\ud V_{\bx'} \right) ^{-1}\,\cdot \notag\\
&= \int_{\bx' \in H_{\bx}} \omega(\bxi )\,{\mathbf v}
\langle \bxi \rangle \otimes \bxi \;\ud V_{\bx'} \cdot
 \left(\int_{\bx' \in H_{\bx}} \omega(\bxi)\,\by
\langle \bxi \rangle \otimes \bxi \;\ud V_{\bx'}\right)^{-1}
\end{align}
where $\bxi=\bx'-\bx $, $\mathbf v \langle \bxi \rangle=\mathbf v(\bx',t)-\mathbf v(\bx,t) $ and $\by \langle \bxi \rangle=\by(\bx',t)-\by(\bx,t)$. It can been seen that the velocity gradient does not require shape tensor, which means the shape tensor is not necessary to define velocity gradient. The incremental deformation gradient can also be obtained without shape tensor $\mathbf K$.

Let $\triangle \bu$ denotes the displacement increment with respect to the previous time step. The incremental spatial deformation gradient $\mathbf C_{\bx} $ in continuum mechanics is defined as
\begin{align}
\mathbf C_{\bx}:=\frac{\partial{(\triangle \bu)}}{\partial{\by}}
=\frac{\partial{(\triangle \bu)}}{\partial{\bx}}\cdot\frac{\partial{\bx}}{\partial{\by}}
=\frac{\partial{(\triangle \bu)}}{\partial{\bx}}\cdot \mathbf F_{\bx}^{-1} \ ,
\end{align}
The non-local counterpart is given by
\begin{align}
\mathbf C_{\bx}:= \int_{\bx' \in H_{\bx}} \omega(\bxi ) \triangle \bu
\langle \bxi \rangle \otimes \bxi \;\ud V_{\bx'} \cdot
 (\int_{\bx' \in H_{\bx}} \omega(\bxi)\by
\langle \bxi \rangle \otimes \bxi \;\ud V_{\bx'})^{-1}\notag \quad \ .
\end{align}
Many material models in non-ordinary state based peridynamic accounting for the geometrical and material non-linear problems are solved on the incremental spatial deformation gradient. The non-ordinary bond force is computed based on the Piola-Kirchhoff as
\begin{align}
\label{eq:Tforce}
\underline{\mathbf T}_{\bx' \bx}=\omega(\bxi)\,\mathbf P_{\bx} \cdot \mathbf K_{\bx}^{-1} \cdot \bxi
\end{align}	
\begin{align}
\label{eq:PK1}
\mathbf P_{\bx}=J_{\bx} \sigma_{\bx} \mathbf F_{\bx}^{-T}, J_{\bx}=\det{\mathbf F_{\bx}}
\end{align}
where $\bxi=\bx'-\bx $,$\sigma_{\bx} $ is Cauchy stress tensor.

Let us define the deformation gradient $\mathbf F_{\bx}'$
\begin{align}
\label{eq:deformationNoShape}
\mathbf F_{\bx}':=\mathbf F_{\bx} \cdot \mathbf K_{\bx} =\int_{\bx' \in H_{\bx}} \omega(\bxi)\by
\langle \bxi \rangle \otimes \bxi \;\ud V_{\bx'}\,\cdot \mathbf K_{\bx}^{-1} \cdot \mathbf K_{\bx}=\int_{\bx' \in H_{\bx}} \omega(\bxi)\by
\langle \bxi \rangle \otimes \bxi \;\ud V_{\bx} \ .
\end{align}
Therefore
\begin{align}
\label{eq:Jx}
 J_{\bx}=\det{ \mathbf F_{\bx}}=\frac{\det{\mathbf F_{\bx}'}}{\det \mathbf K_{\bx}} \ .
\end{align}
By substituting Eqs. (\ref{eq:Jx}), (\ref{eq:deformationNoShape}) and (\ref{eq:PK1}) to (\ref{eq:Tforce}), we obtain
\begin{align}
\label{eq:TforceNoShape}
 \underline{\mathbf T}_{\bx' \bx}
&=\omega(\bxi)\,J_{\bx} \sigma_{\bx} \mathbf F_{\bx}^{-T} \cdot \mathbf K_{\bx}^{-1} \cdot \bxi \notag\\
&=\omega(\bxi)\,\frac{\det{ \mathbf F_{\bx}'}}{\det{ \mathbf K_{\bx}}}\cdot \sigma_{\bx}\cdot (\mathbf F_{\bx}' \cdot \mathbf K_{\bx}^{-1})^{-T} \cdot \mathbf K_{\bx}^{-1} \cdot \bxi \notag\\
&=\omega(\bxi)\,\frac{\det{\mathbf F_{\bx}'}}{\det{\mathbf K_{\bx}}} \cdot \sigma_{\bx}\cdot \mathbf F_{\bx}'^{-T} \cdot \mathbf K_{\bx} \cdot \mathbf K_{\bx}^{-1} \cdot \bxi \notag\\
&=\omega(\bxi)\,\frac{\det{\mathbf F_{\bx}'}}{\det{ \mathbf K_{\bx}}} \cdot \sigma_{\bx}\cdot \mathbf F_{\bx}'^{-T} \cdot \bxi \ .
\end{align}
The shape tensor is only needed for $J_{\bx}=\det{\mathbf{F}_{\bx}}$, while the computation of the bond force density does not require the shape tensor. For any particle with horizon $H_{\bx}$ and dual-horizon $H'_{\bx}$ , the active force $\bff_{\bx \bx'} $ and the passive force $\bff_{\bx' \bx} $ in motion Eq. (\ref{eq:discreteMotion}) or (\ref{eq:intMotion}) can be calculated by
\begin{align}
\bff_{\bx \bx'}=\underline{\mathbf T}_{\bx \bx'}\qquad \forall \bx' \in H_{\bx}' \ ,
\end{align}
and
\begin{align}
\bff_{\bx' \bx}=\underline{\mathbf T}_{\bx' \bx}\qquad \forall \bx' \in H_{\bx} \ ,
\end{align}
where $\bxi =\bx'-\bx $, and $\bx$ and $\bx' $ are the coordinate vectors for particle $\bx$ and particle $\bx'$ in the material configuration, respectively.

\newsavebox\cBBPD
\begin{lrbox}{\cBBPD}
  \begin{minipage}[c]{0.3\textwidth}
    \begin{align*}
      &\bff_{\bx\bx'}-\bff_{\bx'\bx}=c\cdot s_{\bx \bx'} \cdot \bn,\forall \bx' \in H_{\bx}\\
			&\mbox{where  } \bn=\frac{ \boldeta +\bxi}{\|{ \boldeta+\bxi}\|}
    \end{align*} 
  \end{minipage}
\end{lrbox}
\newsavebox\cOSBPD
\begin{lrbox}{\cOSBPD}
  \begin{minipage}[c]{0.3\textwidth}
    \begin{align*}
      &\bff_{\bx\bx'}=\underline{t} \langle \bxi \rangle \cdot  \bn,\\
			&\bff_{\bx'\bx}=\underline{t}' \langle -\bxi \rangle \cdot (-\bn),\forall \bx' \in H_{\bx}\\
			&\mbox{where  } \bn=\frac{ \boldeta +\bxi}{\|{ \boldeta+\bxi}\|}
    \end{align*} 
  \end{minipage}
\end{lrbox}

\newsavebox\cNOSBPD
\begin{lrbox}{\cNOSBPD}
  \begin{minipage}[c]{0.3\textwidth}
    \begin{align*}
      &\bff_{\bx\bx'}=\underline{\mathbf T}_{\bx \bx'},\\
			&\bff_{\bx'\bx}=\underline{\mathbf T}_{\bx' \bx},\forall \bx' \in H_{\bx}
    \end{align*} 
  \end{minipage}
\end{lrbox}

\newsavebox\dBBPD
\begin{lrbox}{\dBBPD}
  \begin{minipage}[c]{0.3\textwidth}
    \begin{align*}
      &\bff_{\bx\bx'}=C(\delta_{\bx'}) \cdot s_{\bx \bx'} \cdot \bn,\forall \bx' \in H_{\bx}';\\
			&\bff_{\bx'\bx}=C(\delta_{\bx}) \cdot s_{\bx \bx'} \cdot (-\bn),\forall \bx' \in H_{\bx}\\
			&\mbox{where  } \bn=\frac{ \boldeta +\bxi}{\|{ \boldeta+\bxi}\|}
    \end{align*} 
  \end{minipage}
\end{lrbox}
\newsavebox\dOSBPD
\begin{lrbox}{\dOSBPD}
  \begin{minipage}[c]{0.3\textwidth}
    \begin{align*}
      &\bff_{\bx\bx'}=\underline{t} \langle \bxi \rangle \cdot  \bn,\forall \bx' \in H_{\bx}';\\
			&\bff_{\bx'\bx}=\underline{t}' \langle -\bxi \rangle \cdot (-\bn),\forall \bx' \in H_{\bx}\\
			&\mbox{where  } \bn=\frac{ \boldeta +\bxi}{\|{ \boldeta+\bxi}\|}
    \end{align*} 
  \end{minipage}
\end{lrbox}

\newsavebox\dNOSBPD
\begin{lrbox}{\dNOSBPD}
  \begin{minipage}{0.3\textwidth}
    \begin{align*}
      &\bff_{\bx\bx'}=\underline{\mathbf T}_{\bx \bx'},\forall \bx' \in H_{\bx}';\\ 
			&\bff_{\bx'\bx}=\underline{\mathbf T}_{\bx' \bx},\forall \bx' \in H_{\bx}
    \end{align*} 
  \end{minipage}
\end{lrbox}

The summary of three kinds of dual-horizon peridynamics see Table.\ref{tab:PDcompare}.
\begin{table}[h]
\begin{center}
\begin{tabular}{ | c | c | c |}
\hline
 & original Peridynamics  & dual-horizon Peridynamics \\ 
Model  
& $\int_{H_{\bx}} (\bff_{\bx\bx'}-\bff_{\bx'\bx})\ud V_{\bx'}$ 
& $\int_{H_{\bx}'} \bff_{\bx\bx'}\ud V_{\bx'}-\int_{H_{\bx}}\bff_{\bx'\bx}\ud V_{\bx'}$
\\ \hline
BB-PD 
&  \usebox{\cBBPD}
&  \usebox{\dBBPD}
\\ \hline
OSB-PD  
&  \usebox{\cOSBPD}
&  \usebox{\dOSBPD}
\\ \hline
NOSB-PD  
&  \usebox{\cNOSBPD}
&  \usebox{\dNOSBPD}
\\ \hline
\end{tabular}
\caption{Comparison of the original Peridynamics and dual-horizon Peridynamics}\label{tab:PDcompare}
\end{center}
\end{table}

\section{Numerical Examples}\label{sec:NumericalTest}
\subsection{Two-dimensional wave reflection in a rectangular plate}\label{subsec:2-dimensional}
\begin{figure}[htp]
	\centering
		\includegraphics{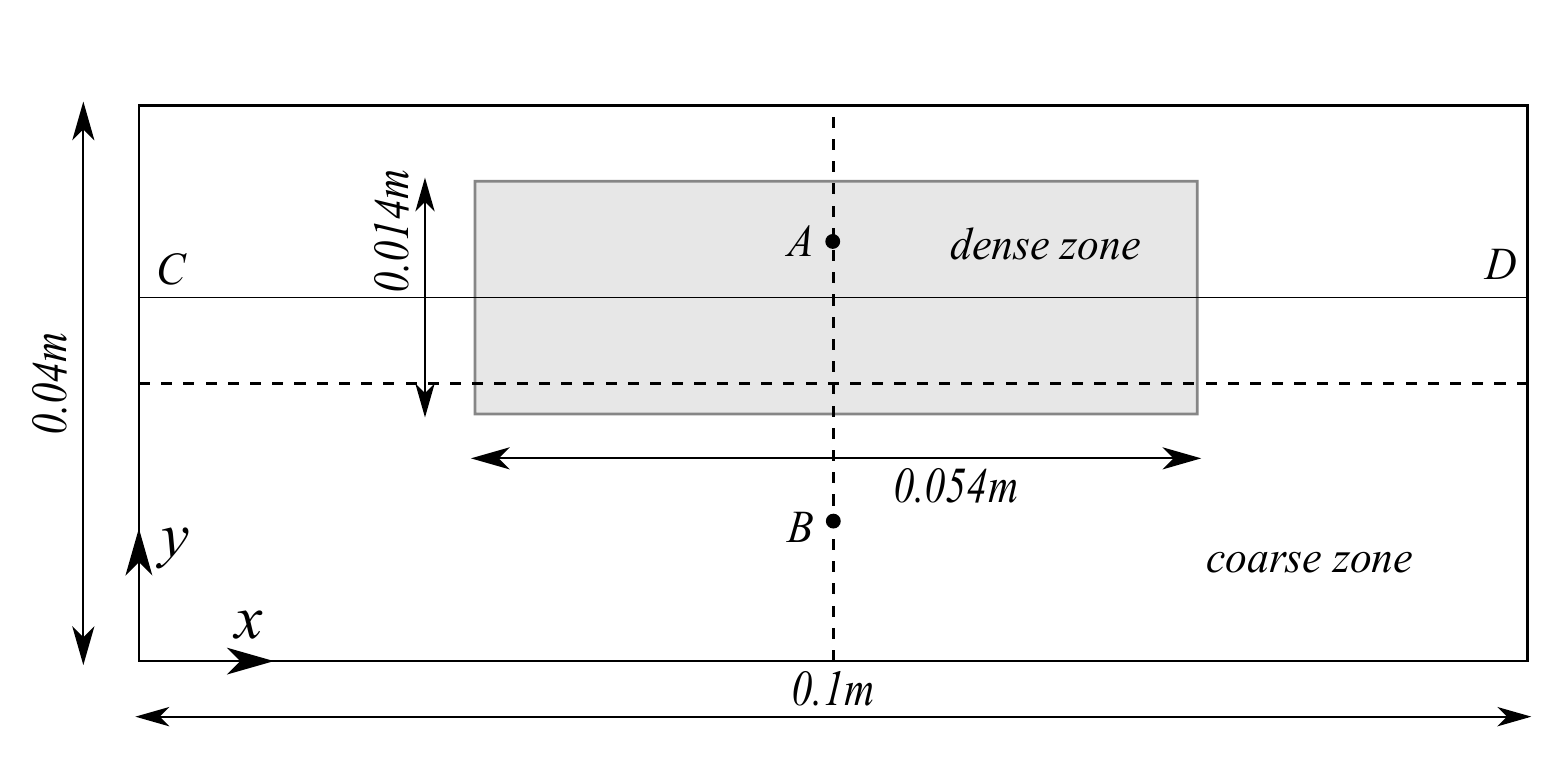}
	\caption{setup of the plate}
	\label{fig:freePlate}
\end{figure}

\begin{figure}[htp]
     \centering
     \subfigure[Particles distribution of model A]{
          \label{fig:MA}
          \includegraphics[width=.6\textwidth]{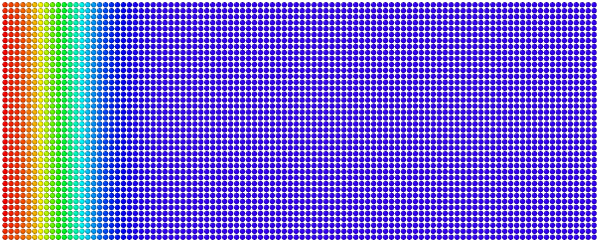}}\\
     \vspace{.1in}
     \subfigure[Particles distribution of model B and C]{
          \label{fig:MBC}
          \includegraphics[width=.6\textwidth]{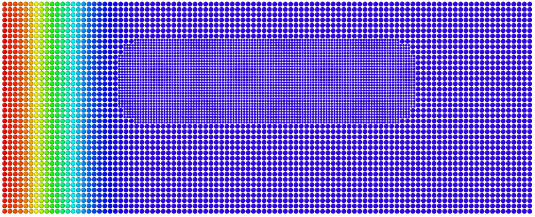}}\\
     \vspace{.3in}
\caption{The discretizations of model A,B and C.}
\label{fig:MABC}
\end{figure}

Consider a rectangular plate with dimensions of 0.1$\times$0.04 $\mbox{m}^2$ (see Fig.\ref{fig:freePlate}). The Young's modulus, density and the Poisson's ratio used for the plate are $E=1$, $\rho=1$ and $\nu=0$, respectively.Note that this is 1-D model. The initial displacement applied to the plate is described by the following equation
\begin{align}
\label{eq:initialDisp}
u_0(x,y)=0.0002 \exp [-(\frac{x}{0.01})^2],\quad v_0(x,y)=0, \quad x \in [0,0.1],y \in [-0.02,0.02] \ ,
\end{align}
where $u_0$ and $v_0$ denote the displacement in the $x$ and $y$ directions respectively. The wave speed is $v=\sqrt{E/\rho}=1 \,\mbox{m/s}$. At any time step, the $L_2$ error in the displacement is given by
\begin{align}
\label{eq:L2}
{\|\text{err}\|}_{L_2}=\frac{\|\mathbf u^h-\mathbf u_{\text{analytic}}\|}{\|\mathbf u_{\text{analytic}}\|} \ ,
\end{align}
with
\begin{align}
\|\mathbf u\|= \left( \int_{\Omega_0} \mathbf u \cdot \mathbf u\, d\Omega_0 \right) ^{\frac{1}{2}} \notag \ .
\end{align}
Three models for solving this problem, namely model A, B and C, are devised,see Figs.\ref{fig:MA} and \ref{fig:MBC}. In model A, the plate is discretised with 4000 particles and all particles have the same horizon sizes of 0.001 m, see Fig.\ref{fig:MA}. For models B and C, we adopted a discretisation of varying horizons as shown in Fig.\ref{fig:MBC}. Verlet integration method is adopted.The horizon radius associated to each particle is set as 3 times the particle sizes for all models.  Therefore, along the interface between the coarse and fine discretization, the horizon sizes vary. The minimal particle size of 3 models is $\triangle x=$5e-4m, yielding a critical time increment $\triangle t_{max}=\triangle x/v$=5e-4s.  The physical computational time step is set 0.5 second with the time increment $\triangle t_{max}$=5e-4s.

As BB-PD can not simulate the model with Poisson ratio $\nu=0$, the OSB-PD model is adopted. Model A was solved with the OSB-PD with constant horizons. Model B is the original OSB-PD with variable horizon (without additional treatment for ghost force), and model C is our dual-horizon formulation for OSB-PD with variable horizon.  The parameters used for the three models are listed in Table.\ref{tab:modelABC}.

\begin{table}[h]
\begin{center}
\begin{tabular}{ | l | r | r | r | }
\hline
Model & $\triangle x= \triangle y $ & $10^3 \cdot \delta$ & Particle numbers \\ \hline
A  & 0.001 & 3  & 4000\\ \hline
B  & 0.001,0.0005 & 3/1.5  & 8628 \\ \hline
C  & 0.001,0.0005 & 3/1.5  & 8628 \\ \hline
\end{tabular}
\caption{Values of the Peridynamic parameters for the mixed model}\label{tab:modelABC}
\end{center}
\end{table}

\begin{figure}[htp]
	\centering
		\includegraphics{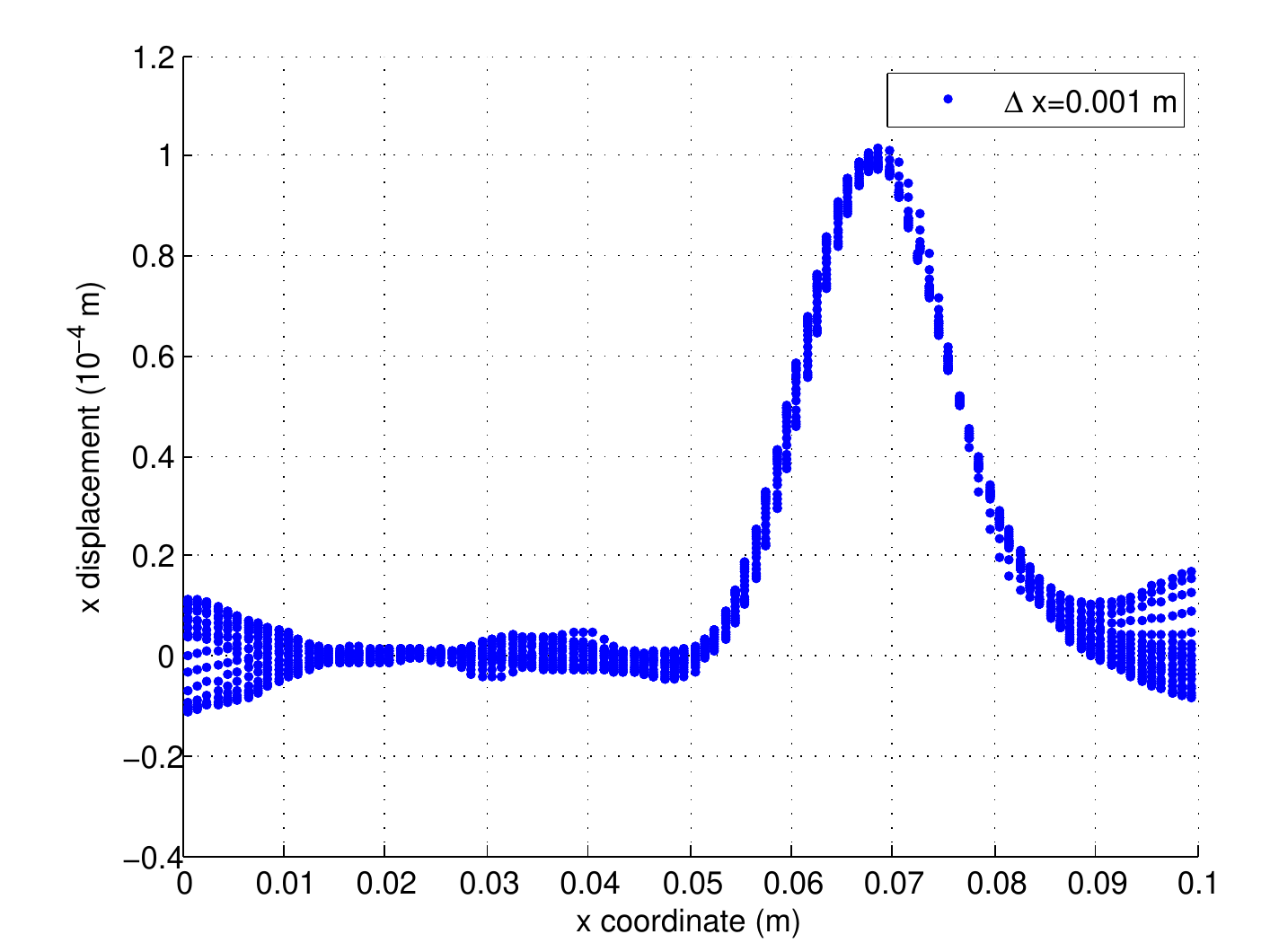}
	\caption{Displacement wave $u_x$ versus $x$ coordinate of model A at step 650}
	\label{fig:AVibrationWave}
\end{figure}

\begin{figure}[htp]
	\centering
		\includegraphics{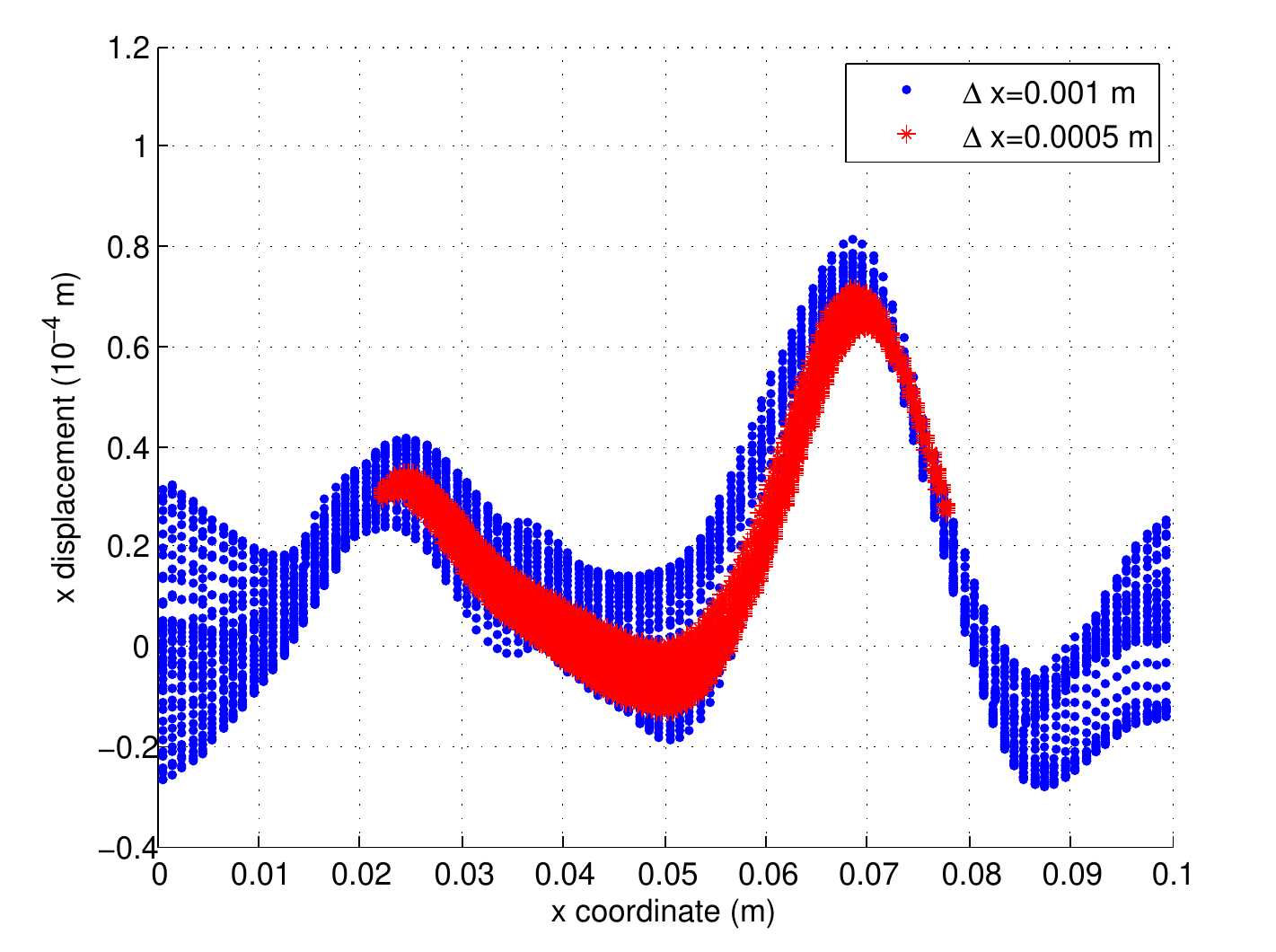}
	\caption{Displacement wave $u_x$ versus $x$ coordinate of model B at step 650}
	\label{fig:BVibrationWave}
\end{figure}

\begin{figure}[htp]
	\centering
		\includegraphics{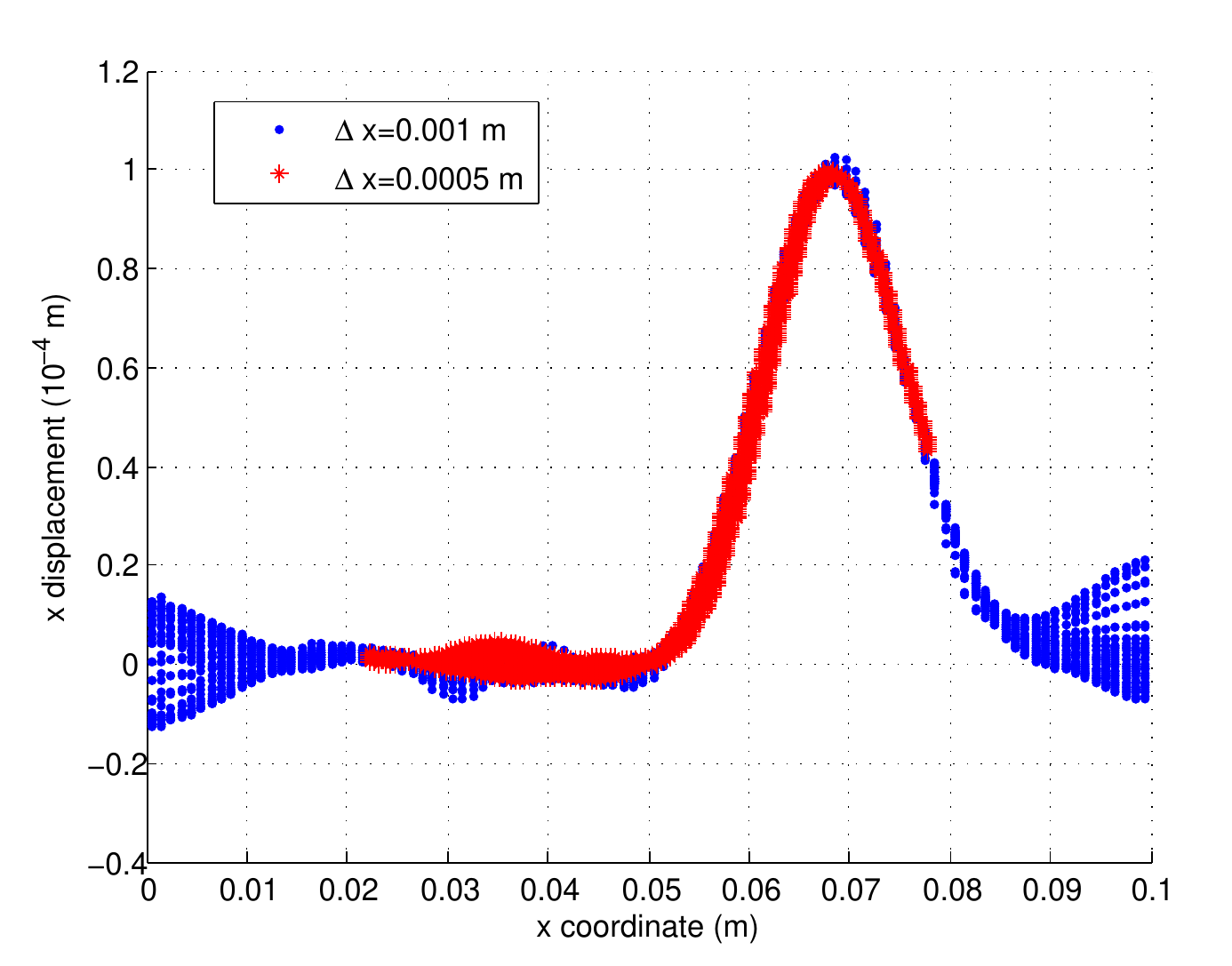}
	\caption{Displacement wave $u_x$ versus $x$ coordinate of model C at step 650}
	\label{fig:CVibrationWave}
\end{figure}

Fig.\ref{fig:AVibrationWave},\ref{fig:BVibrationWave},\ref{fig:CVibrationWave} show the displacement in the x-direction along the x-coordinate of all particles. The red star points are  particles with $\Delta x=0.0005 \mbox{m}$, the blue dot particles refer to $\Delta x=0.001 \mbox{m}$.
The displacement in model C (present dual-horizon PD) is almost identical to that of model A (PD with constant horizon),as shown in Fig.\ref{fig:AVibrationWave} and Fig.\ref{fig:CVibrationWave}. Spurious wave reflections are observed for Model B, as shown in Fig.\ref{fig:BVibrationWave}. At step 650, several wave peaks are observed in Model B and the maximum displacement is lower than that of Models A and C. Therefore, the displacement wave in Model B were affected by spurious wave reflection and the results deviate greatly from results of A and C.

We compute the $L_2$ errors of the wave profile along the line-$CD$ (see Fig. \ref{fig:freePlate}). The $L_2$ error (see Table. \ref{tab:L2error}) of the present peridynamic formulation, can achieve the almost the same accuracy as the conventional peridynamics with constant horizon.

\begin{figure}[htp]
     \centering
     \vspace{-.1in}
		 \subfigure[$x$ displacement of model C]{
          \label{fig:Cux}
          \includegraphics[width=.44\textwidth]{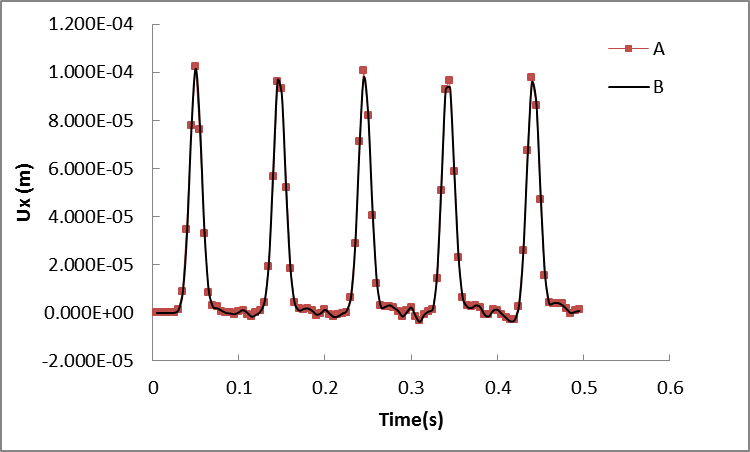}}
     \hspace{-.1in}
      \vspace{-.1in}
			\subfigure[$y$ displacement of model C]{
     \label{fig:Cuy}
          \includegraphics[width=.44\textwidth]{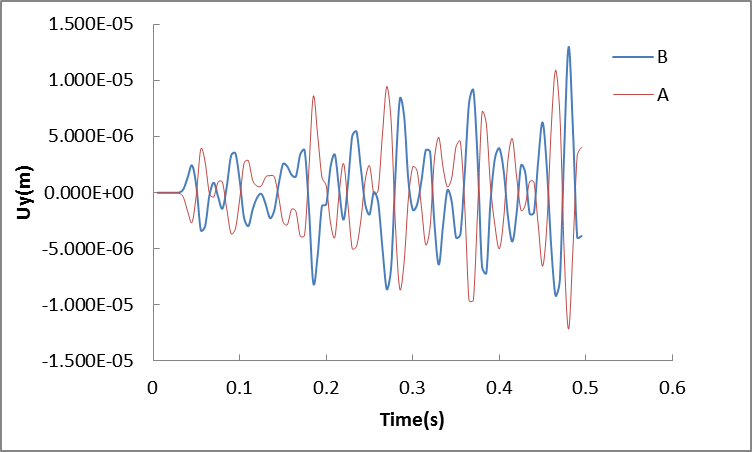}}\\
     \subfigure[$x$ velocity of model C]{
     \label{fig:Cvx}
     \includegraphics[width=.44\textwidth]{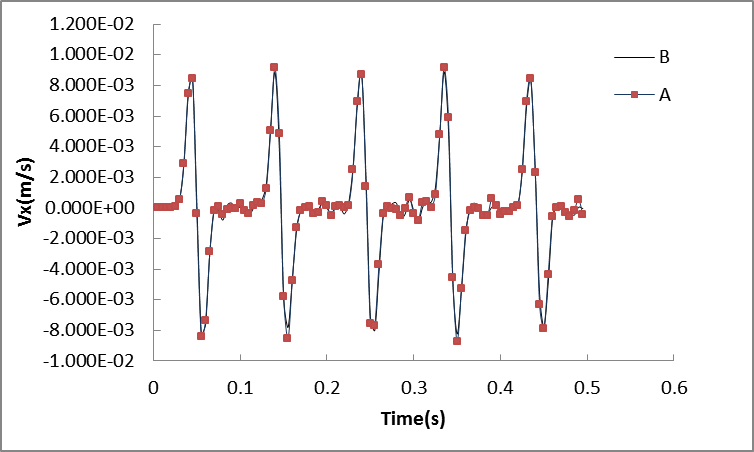}}
     \hspace{-.1in}
     \subfigure[$y$ velocity of model C]{
     \label{fig:Cvy}
     \includegraphics[width=.44\textwidth]{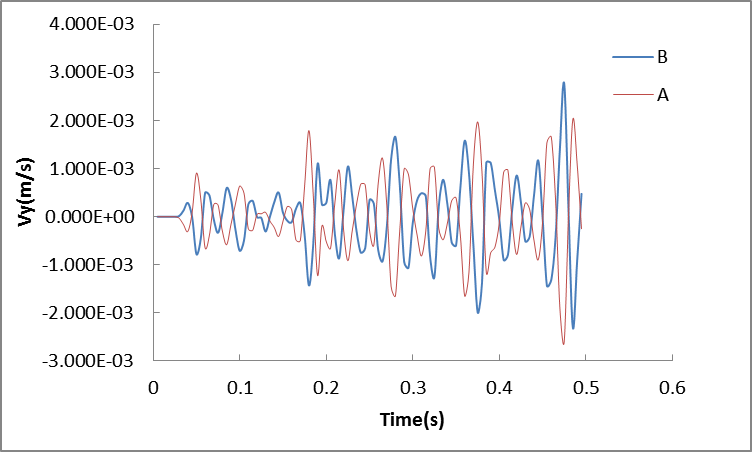}}\\
		
		\vspace{-.1in}
		 \subfigure[$x$ displacement of model B]{
          \label{fig:Bux}
          \includegraphics[width=.44\textwidth]{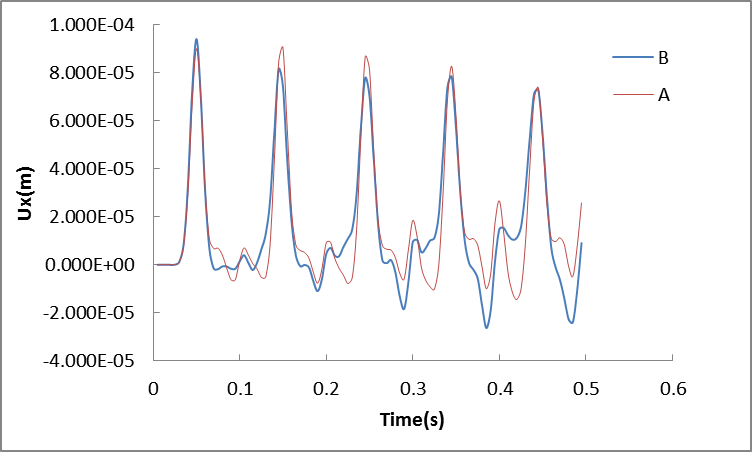}}
     \hspace{-.1in}
      \vspace{-.1in}
			\subfigure[$y$ displacement of model B]{
     \label{fig:Buy}
          \includegraphics[width=.44\textwidth]{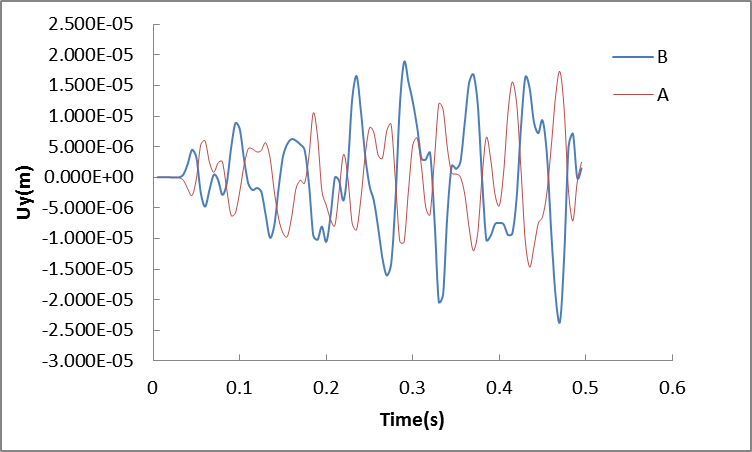}}\\
     \subfigure[$x$ velocity of model B]{
     \label{fig:Bvx}
     \includegraphics[width=.44\textwidth]{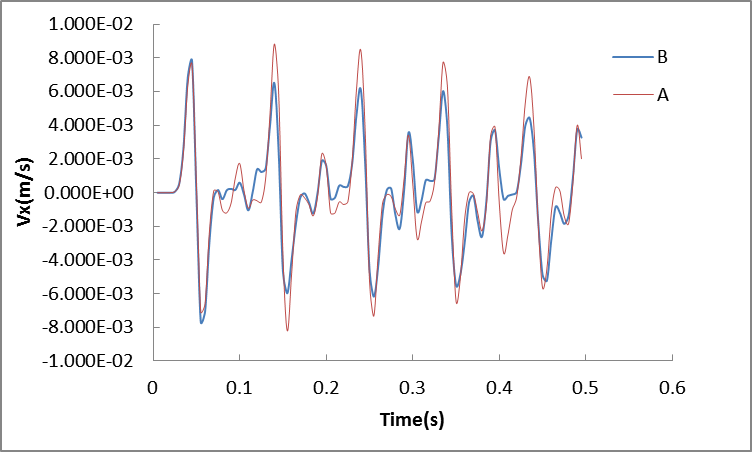}}
     \hspace{-.1in}
     \subfigure[$y$ velocity of model B]{
     \label{fig:Bvy}
     \includegraphics[width=.44\textwidth]{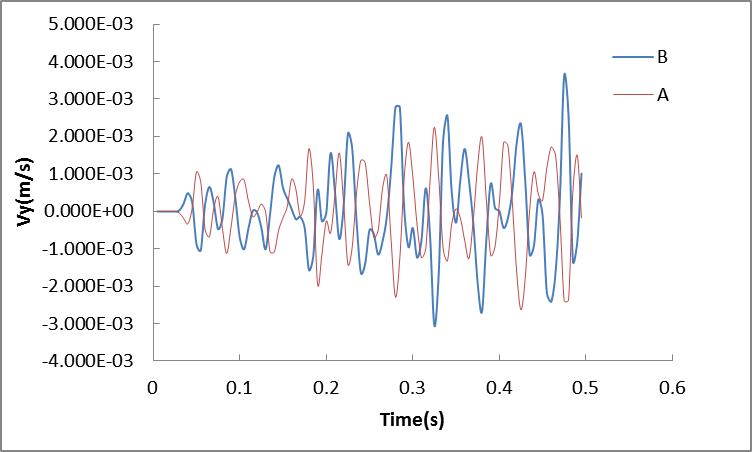}}
\caption{The displacement and velocity curve of monitor points $A$ and $B$ for model B and C}
\label{fig:MonitorPointCurve}
\end{figure}

\begin{table}[h]
\begin{center}
\begin{tabular}{ | l | r | r | r | r |}
\hline
Model & step 100 & step 200 & step 650 & step 990\\ \hline
A  & 0.0410 & 0.0452  & 0.1277 & 0.2498 \\ \hline
B  & 0.1149 & 0.1738  & 0.7638 & 1.3233 \\ \hline
C  & 0.0447 & 0.0500  & 0.1253 & 0.2871 \\ \hline
\end{tabular}
\caption{$L_2$ error of displacement for the initial condition (Gauss distribution of displacement) before and after the wave reflection }\label{tab:L2error}
\end{center}
\end{table}
\noindent \textbf{Displacement and velocity of monitor points}\\
Two monitor points (A,B) are selected to evaluate the spurious wave reflection of Models B and C as shown in Fig. \ref{fig:freePlate}. The displacement and velocity curves show the spurious wave reflection exists in model B while not exists in model C, see Fig.\ref{fig:MonitorPointCurve}.

\subsection{Kalthof-Winkler experiment}\label{subsec:3DWinkler}

\begin{figure}[htp]
	\centering
		\includegraphics[width=12cm]{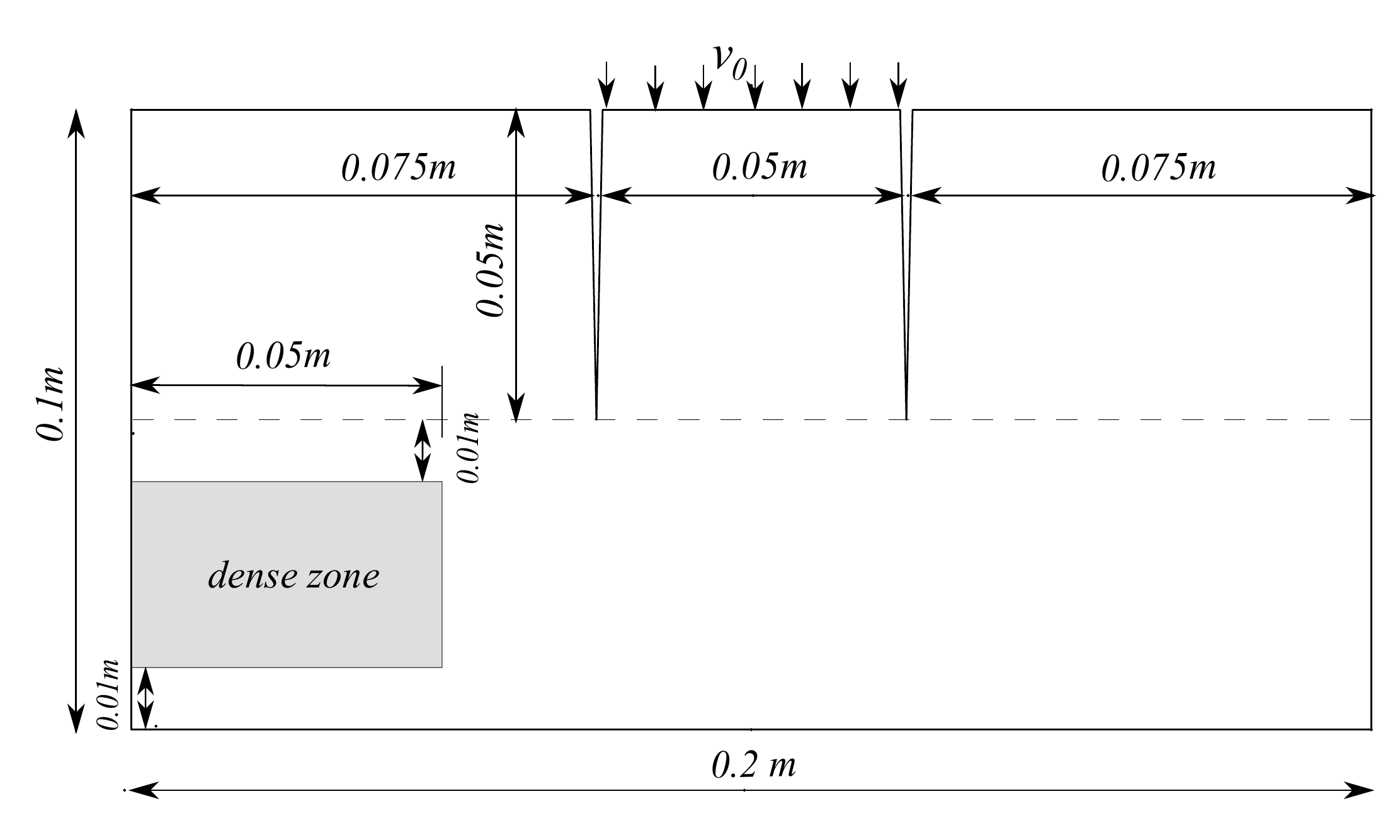}
	\caption{Kalthof-Winkler's experimental setup}
	\label{fig:winkler2}
\end{figure}

\begin{figure}[htp]
	\centering
		\includegraphics[width=9cm]{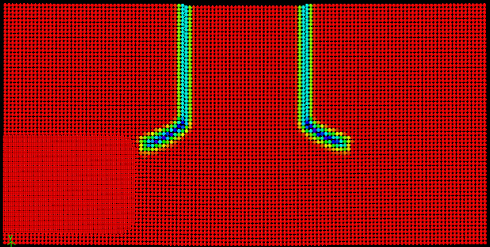}
	\caption{The crack pattern of Kalthof-Winkler plate by the present dual horizon PD at step 350}
	\label{fig:CPHVWinkler350}
\end{figure}

\begin{figure}[htp]
	\centering
		\includegraphics[width=9cm]{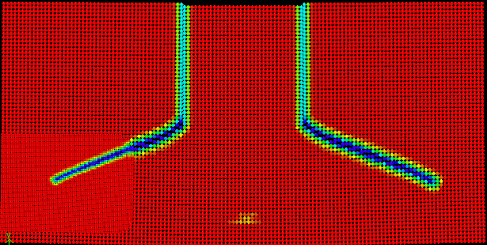}
	\caption{The crack pattern of Kalthof-Winkler simulation by the present dual horizon PD at step 650}
	\label{fig:CPHVWinkler650}
\end{figure}

\begin{figure}[htp]
	\centering
		\includegraphics[width=9cm]{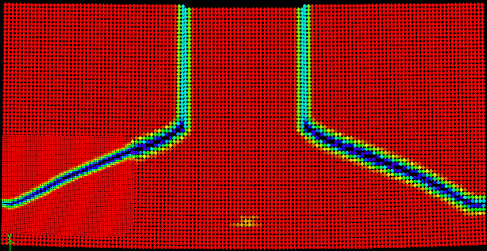}
	\caption{The crack pattern of Kalthof-Winkler simulation by the present dual horizon PD at step 875}
	\label{fig:CPHVWinkler875}
\end{figure}

\begin{figure}[htp]
	\centering
		\includegraphics[width=9cm]{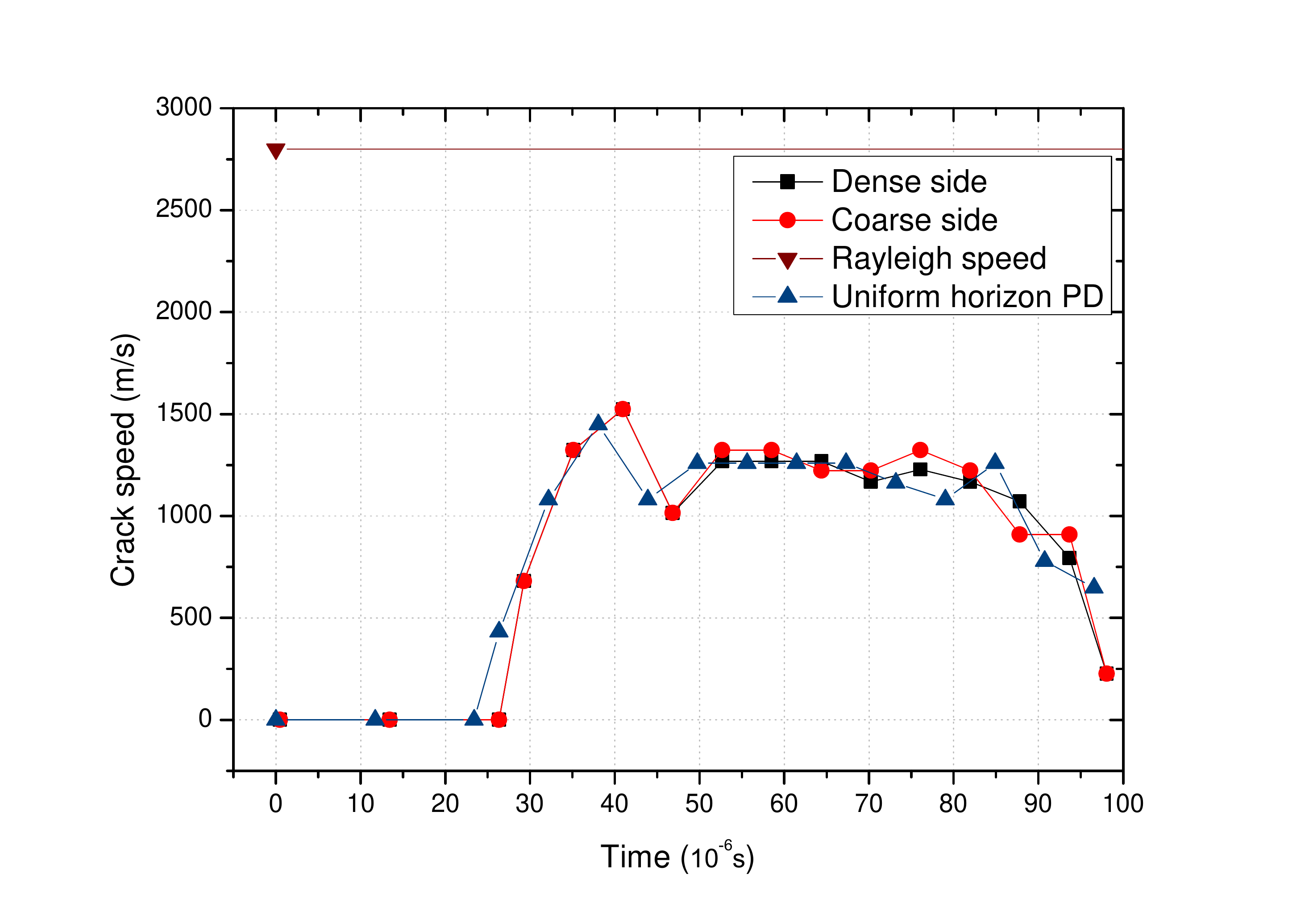}
	\caption{The crack speed of Kalthof-Winkler simulation by the present dual horizon PD}
	\label{fig:CPHVWinkler}
\end{figure}

To study the performance of the present formulation for fracture modeling, the Kalthof-Winkler experiment  is exploited. The geometry and model used for the test is depicted in Fig.\ref{fig:winkler2}, the thickness of the specimen is set as 0.01m. In the experiment, the evolution of crack pattern was observed to be dependent on the impact loading velocity. For a plate made of steel 18Ni1900 subjected to an impact loading at the speed of $v_0$=32 m/s, brittle fracture was observed \cite{Silling2001}. The crack propagates from the end of the initial crack at an angle around 70$^{\circ}$ vs. the initial crack direction. In \cite{Silling2001}, the BB-PD with constant horizon was used to test this example. For convenience of comparison, the present dual-horizon formulation is also applied to the bond-based peridynamics to test the example. The material parameters used are the same as in \cite{Silling2001}, i.e. the elastic modulus $E=190 \,\mbox{GPa}$,$\rho=7800\, \mbox{kg/m}^3$,$\nu=0.25$ and the energy release rate $G_0=6.9e4\,\mbox{J/m}^2$. The impact loading was imposed by applying an initial velocity at $v_0=22\, \mbox{m/s}$ to the first three layers of particles in the domain as shown in Fig.\ref{fig:winkler2};all other boundaries are free. The plate is discretized with two different particle sizes, namely $\triangle x_{\text{coarse}}=1.5625\mbox{e-}3\,\mbox{m}$ for the coarse subdomain and $\triangle x_{\text{dense}}=0.5\triangle x_{\text{coarse}}=7.8125\mbox{e-}4\,\mbox{m}$ for the fine subdomain located in the left down corner of the model, see Fig.\ref{fig:winkler2}. Along the thickness of the plate (in direction perpendicular to the plane surface), four layers of particles in the coarse subdomain and eight layers in the fine subdomain are employed. The total number of particles is 57968.
The crack propagation speed is computed by
\begin{align}
V_{l-0.5}=\frac{\|\mathbf x_l-\mathbf x_{l-1}\|}{t_{l}-t_{l-1}} \ ,
\end{align}
where $ \mathbf x_l$ and $\mathbf x_{l-1}$ are the positions of the crack tip at the times $t_l$ and $t_{l-1}$ respectively. An expression for the Rayleigh wave speed $c_R$ \cite{Graff1975} is given as
\begin{align}
\frac{c_R}{c_s}=\frac{0.87+1.12\nu}{1+\nu}\label{eq:RayleighSpeed} \ ,
\end{align}
where $\nu$ is the Poisson's ration, $c_s=\sqrt{\mu/\rho}$ is the shear wave speed and $\mu$ is the shear modulus. The crack starts to propagate at 26.3 $\mu$s. The highest crack speed reached is 1530 m/s, about 54.4\% of the Rayleigh speed(2799.2 m/s). The average crack speed is 1077 m/s for the fine subdomain and 1094 m/s for the coarse subdomain. During the crack propagation, the crack paths in the fine and coarse subdomains are nearly symmetrical, as shown in Figs. \ref{fig:CPHVWinkler350}, \ref{fig:CPHVWinkler650} and \ref{fig:CPHVWinkler875}. It can be seen from  Fig.\ref{fig:CPHVWinkler} that the crack speed is very close to that predicted by the PD formulation using constant horizons ($\triangle x_{\text{uniform}}=1.5625e\mbox{-3\,m}$). The crack propagation initiated at an angle of 65.7$^{\circ}$ in the fine subdomain with respect to the original crack and 65.8$^{\circ}$ in fine subdomain. Therefore, it can be concluded that with the present formulation, the transition from fine to coarse of horizons has almost no influence on the crack propagation speed.

\subsection{Adaptively refined peridynamics}
\begin{figure}[htp]
	\centering
		\includegraphics[width=9cm]{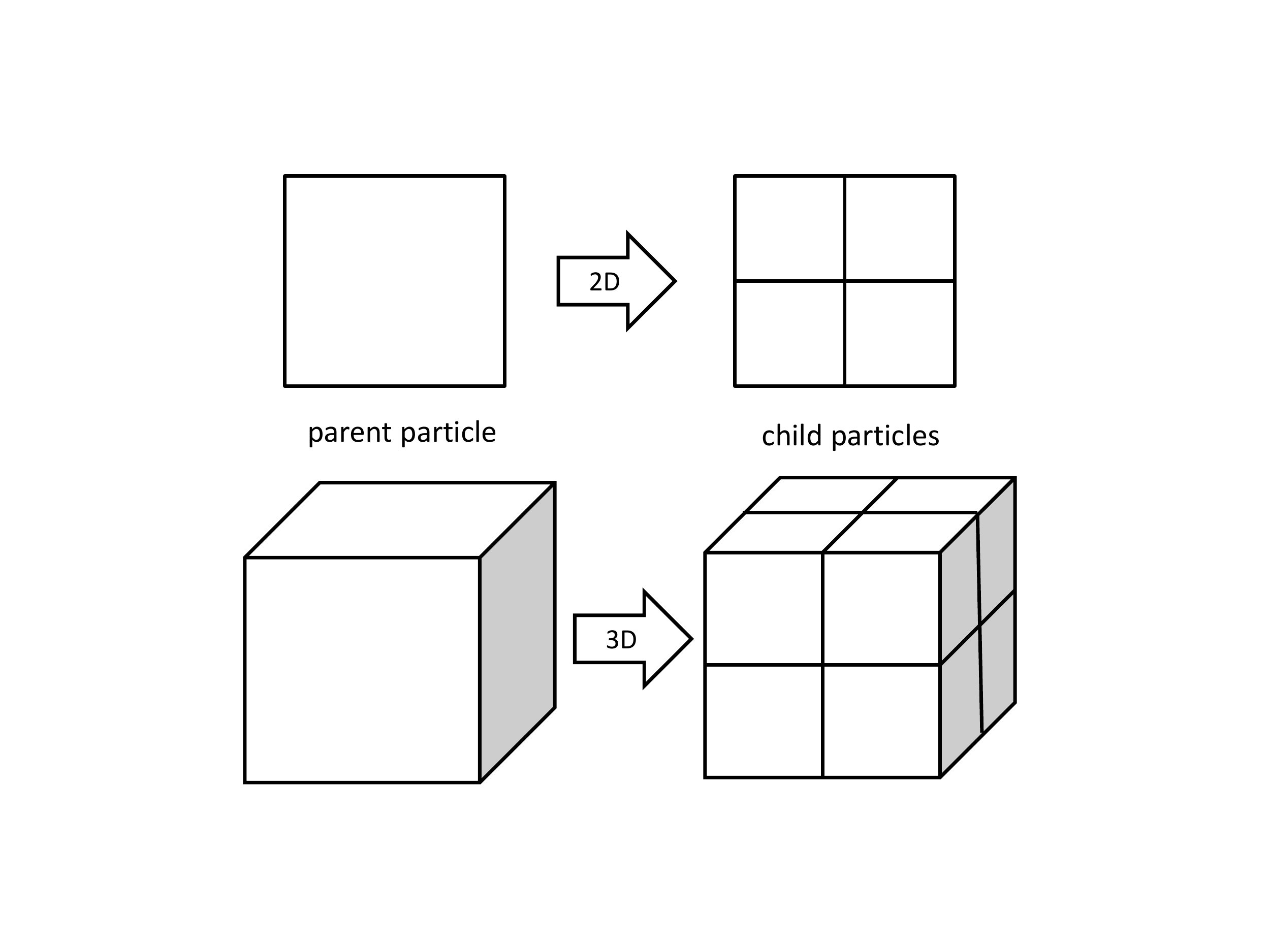}
	\caption{Particle splitting for Adaptive refined peridynamics}
	\label{fig:refineSplit}
\end{figure}

The present dual-horizon Peridynamic formulation provides the possibility for adaptive-refinement within a simple and unified framework. Two examples of adaptively refined peridynamics will be tested in this section: the Kalthof Winkler experiment in section \ref{subsec:3DWinkler} and plate with pre-crack subjected to traction. The threshold for the adaptive refinement is determined by both the damage-state criteria and energy state, as proposed in \cite{Dipasquale2014}. The adaptive refinement procedure consists of two steps.\\
(1) Search for the particles that exceed the threshhold values. Note that the refinement is not only applied to the particles above the threshold but also to the neighbouring particles. In this way, it can be ensure that the crack tip always remains inside the refined zone. \\
(2) Split the particle, named as parent particle, into small particles, named as child particles. The properties that will be mapped from the parent particles to the child particles include the mass, volume, coordinate, displacement and velocity. \\
The method to split particles for structured discretization is shown in Fig. \ref{fig:refineSplit}. Splitting the particle results in halving the maximum time step. In all examples, we restrict each particle being allowed to  split only once in the entire analysis.

\subsubsection{3D adaptively refined Kalthof-Winkler simulation}\label{sssec:adaptiveWinkler}
We now repeat the Kalthof Winkler experiment from section \ref{subsec:3DWinkler}. However, the initial discretization is based on uniform horizon size. The initial particle size is $\triangle x_{\text{initial}}=$1.5625e-3 m. After refinement, the particle size is reduced to $\triangle x_{\text{refined}}=$7.8125e-4 m around the crack. The total particles is increased from 32,768 to 54,860 at the end of the simulation. A uniform refinement would result 32,768$\times$8=262,144 partilces.

\begin{figure}[htp]
	\centering
		\includegraphics[width=9cm]{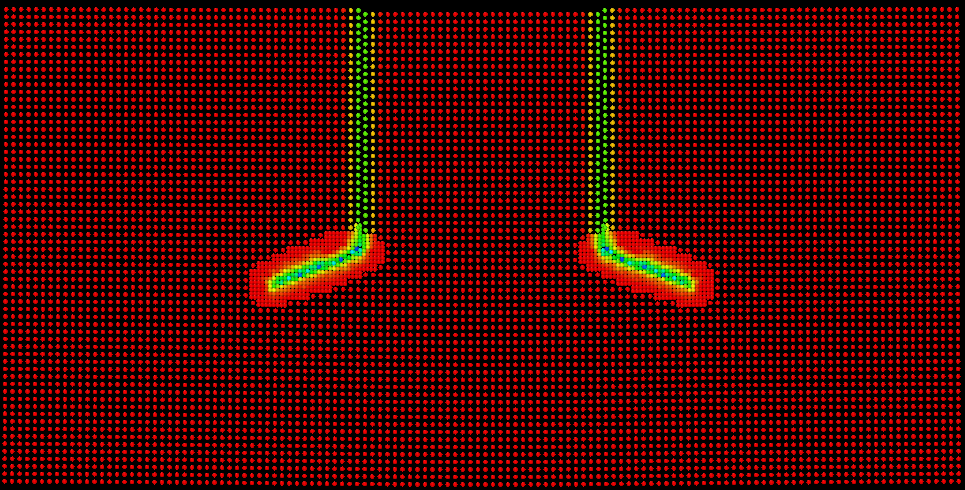}
	\caption{The crack pattern of adaptive refined Kalthof-Winkler experiment at step 420}
	\label{fig:CPadaptiveWinkler420}
\end{figure}

\begin{figure}[htp]
	\centering
		\includegraphics[width=9cm]{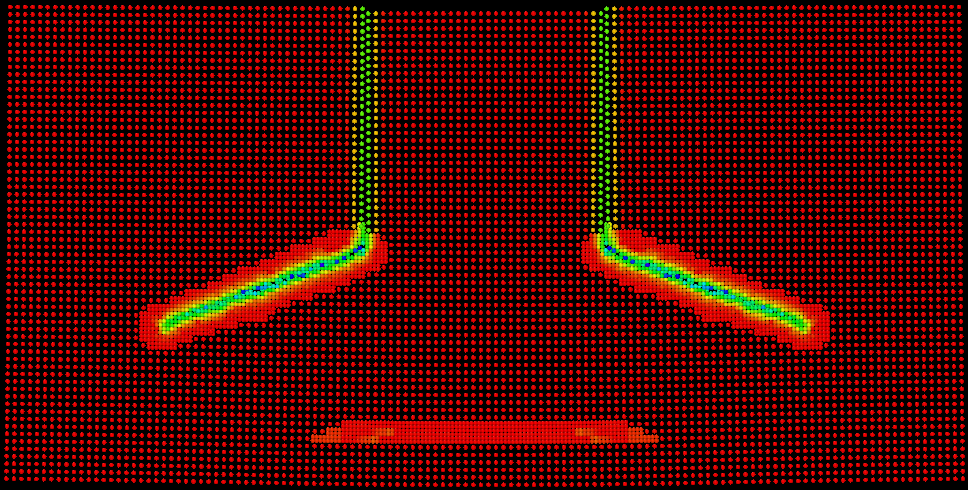}
	\caption{The crack pattern of adaptive refined Kalthof-Winkler experiment at step 600}
	\label{fig:CPadaptiveWinkler600}
\end{figure}

\begin{figure}[htp]
	\centering
		\includegraphics[width=9cm]{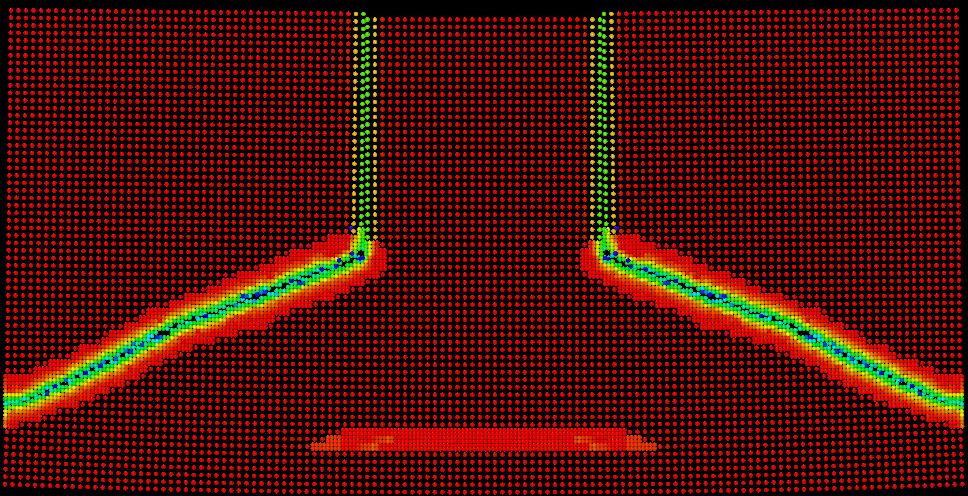}
	\caption{The crack pattern of adaptive refined Kalthof-Winkler experiment at step 925}
	\label{fig:CPadaptiveWinkler925}
\end{figure}

\begin{figure}[htp]
	\centering
		\includegraphics[width=9cm]{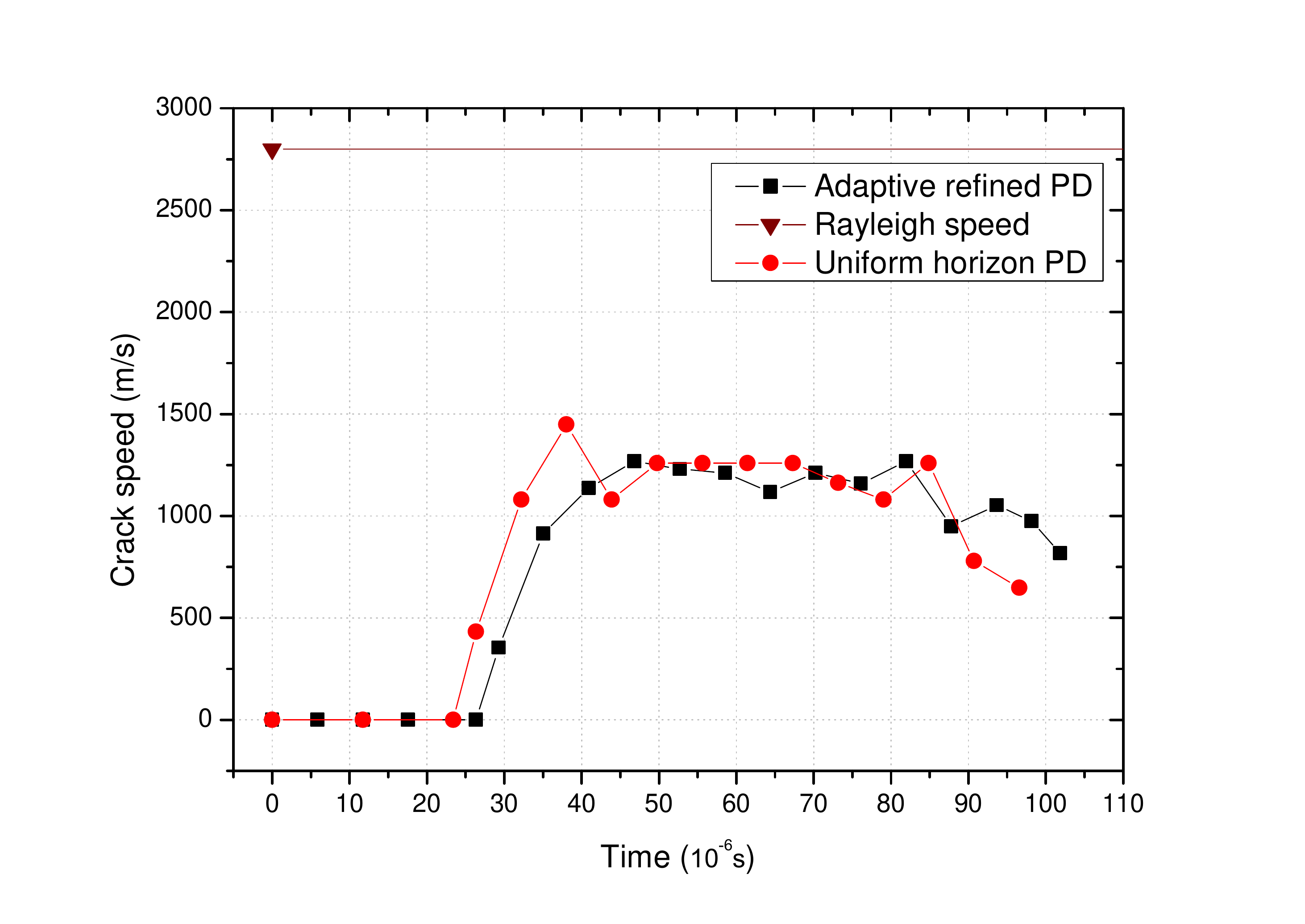}
	\caption{The crack speed of adaptive refined Kalthof-Winkler experiment}
	\label{fig:CPadaptiveWinkler}
\end{figure}

The crack patterns at certain steps are plotted in Figs.\ref{fig:CPadaptiveWinkler420}, \ref{fig:CPadaptiveWinkler600} and \ref{fig:CPadaptiveWinkler925}. The crack tip is always contained inside the refined zone. According to Fig. \ref{fig:CPadaptiveWinkler}, the maximum crack propagation speed reaches 1268 m/s and the average speed is 1047.6 m/s. The crack propagation takes place at an angle of approximately 66.5$^{\circ}$ with respect to the initial crack direction.

\subsubsection{Plate with pre-crack subjected to traction}\label{sssec:adaptivePlate}
\begin{figure}[htp]
	\centering
		\includegraphics[width=9cm]{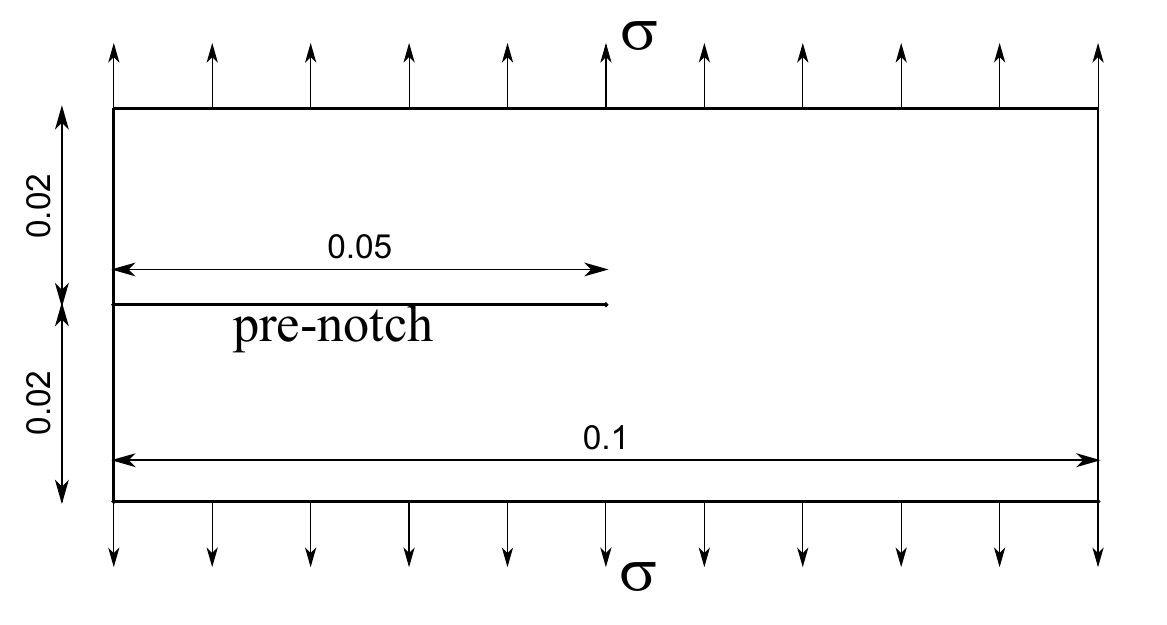}
	\caption{Setup of the pre-cracked plate under traction load}
	\label{fig:platesetup}
\end{figure}

In the last example, we will test the capability of the new formulation in modelling the crack branching. Hence, a plate with pre-crack subject to traction is considered as shown in Fig.\ref{fig:platesetup}. The traction remains constant during the entire simulation. The same example has been well studied in \cite{Ha2010} with BB-PD and also in \cite{Belytschko2003} using the XFEM.

The material parameters of the plate (soda-lime glass) are $E=72 \,\mbox{GPa}$, $\rho=2440\, \mbox{kg/m}^3$ and energy release rate being $G_0=135\,\mbox{J/m}^2$ \cite{Ha2010}. Two models, namely Case 1 and Case 2, are set up for comparison. Case 1 is solved by using the traditional BB-PD with constant horizons, while Case 2 is modeled by an adaptively refined BB-PD using the present dual-horizon formulation. The plate is assumed under plane stress condition and all simulations are carried out in 2D. The particle sizes chosen are $5\times10^{-4}$ m for Case 1 and $1\times10^{-3}$ m for Case 2, respectively. Case 2 uses an adaptively refined model, and once a particle is refined, the minimal particle size will become the same as in Case 1. The particle number are 16,000 for Case 1 and 4000-6424 for Case 2, respectively. The Rayleigh speed for the material is 3102 m/s according to Eq. (\ref{eq:RayleighSpeed}). The max crack speed is 1881.4 m/s (60.6\% of $c_R$) for Case 1 and 2184.1 m/s (70.4\% of $c_R$) for Case 2, respectively. One possible reason for the different max crack speeds is the deviation of the mapping method.

For Case 1, the crack starts to propagate at 11.9 $\mu$s, and the first crack branching point $B1$ occurs at 24.5 $\mu$s at the speed of 1043.6 m/s and the second crack branch point $B2$ takes place at 40.9 $\mu$s at the speed of 1147.1 m/s. It is observed once branching initiates, the crack propagation speed decreases. This can be explained that the energy release to create more fracture surfaces decelerates the propagation speed. Afterward, the crack speed increases. For Case 2, the initial crack started to propagate at 10.3 $\mu$s, first branching point $B1$ at 19.4 $\mu$s with 1247.6 m/s, and second branching point $B2$ at 34.2 $\mu$s with 1330.6 m/s.  The  crack pattern is similar compared with Case 1.

\begin{figure}[htp]
	\centering
		\includegraphics[width=9cm]{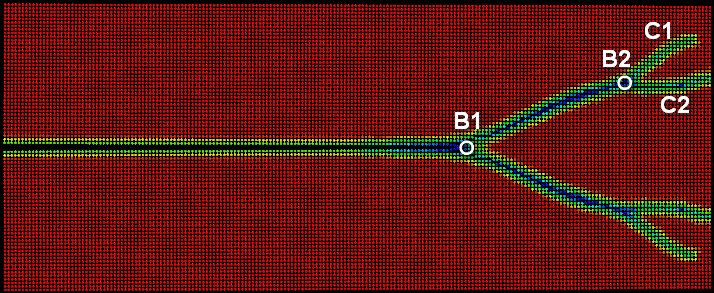}
	\caption{The crack pattern of pre-cracked plate with uniform horizons}
	\label{fig:CPconstplate}
\end{figure}

\begin{figure}[htp]
	\centering
		\includegraphics[width=9cm]{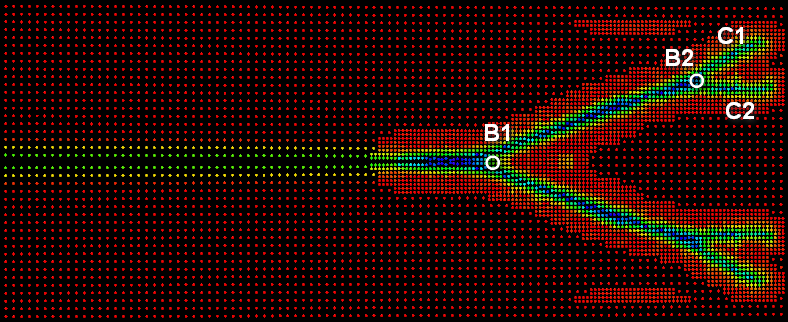}
	\caption{The crack pattern of adaptive refined pre-cracked plate}
	\label{fig:CPadaptiveplate}
\end{figure}

\begin{figure}[htp]
	\centering
		\includegraphics[width=9cm]{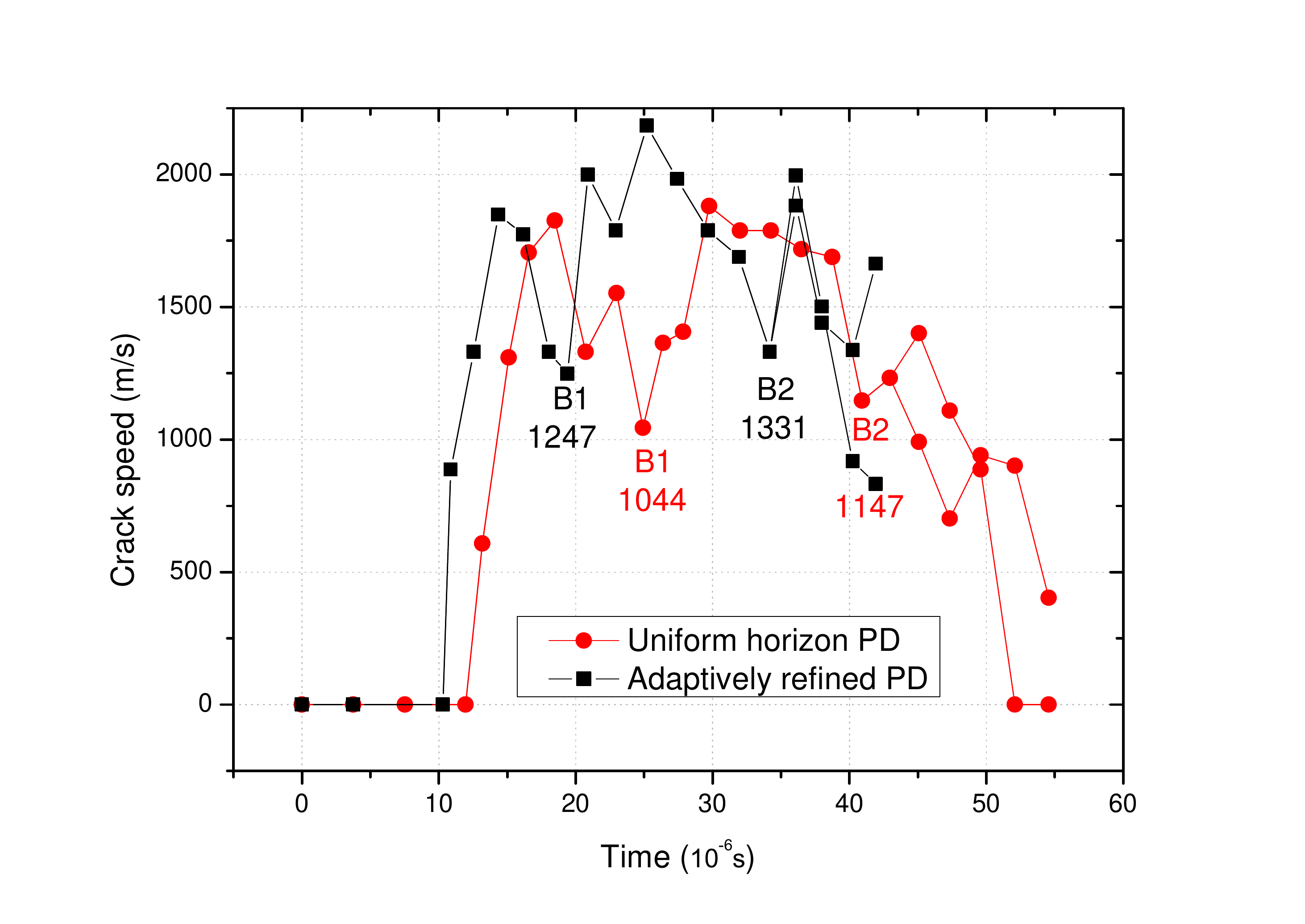}
	\caption{The crack speed of pre-cracked plate under traction}
	\label{fig:CPfineplate}
\end{figure}

\section{Conclusions}\label{sec:Conclusions}
This paper contributes to the development of peridynamics with varying horizons. Therefore, the interactions between particles are based on two independent horizons, passive force from the horizon and the active force from the dual-horizon. The spurious wave reflection and ghost force in the conventional peridynamics is naturally eliminated. Based on the new concepts of horizons, the motion equation of dual-horizon peridynamics is derived. We show that the balance of linear momentum and angular momentum is satisfied. The key difference of the present peridynamics formulation from the conventional one is the way of computing reactions forces. Hence, the dual-horizon peridynamics can be implemented in any existing peridynamics code with minimal efforts. Three numerical examples show the present method is free from spurious wave reflection, and the accuracy is retained along the interface where horizon sizes undergo sudden changes. The method also shows its capability for fracture problems including crack branching. A simple h-adaptive scheme is proposed to improve crack paths resolution, both for 2D and 3D case. We also intend to develop efficient error estimate for adaptivity to drive the adaptive refinement.  The present method  also shows the potential in multiscale analysis where the models at different length scales can be bridged by using different horizon settings.

\section*{Acknowledgements}
The authors acknowledge the supports from FP7 Marie Curie Actions ITN-INSIST and IIF-HYDROFRAC (623667), the National Basic Research Program of China (973 Program: 2011CB013800) and NSFC (51474157), the Ministry of Science and Technology of China (Grant No.SLDRCE14-B-28, SLDRCE14-B-31). The authors would like to express the gratitude to Dr. Jafar Dashlejeh and Dr. Erkan Oterkus for the beneficial discussion about Peridynamic theory.

\section*{References}
\bibliographystyle{unsrt}
\bibliography{Horizon_PD}

\end{document}